\documentclass[fleqn,usenatbib]{mnras}

\usepackage{newtxtext,newtxmath}
\usepackage[T1]{fontenc}

\DeclareRobustCommand{\VAN}[3]{#2}
\let\VANthebibliography\thebibliography
\def\thebibliography{\DeclareRobustCommand{\VAN}[3]{##3}\VANthebibliography}

\usepackage{graphicx}	% Including figure files
\usepackage{amsmath}	% Advanced maths commands
\usepackage{multirow}
\usepackage[table,xcdraw]{xcolor}
\usepackage{array}
\usepackage{xcolor}
\usepackage{colortbl}
\usepackage{booktabs}
\usepackage{pgf}
\usepackage{tikz}
\usepackage{hhline}
\usepackage{makecell}

\usepackage{xcolor}
\usepackage[textwidth=2.5cm]{todonotes} % optional, for margin notes

\usepackage{xcolor}
\usepackage{colortbl}
\usepackage{booktabs}
\usepackage{orcidlink}

\definecolor{constraintpoor}{RGB}{180, 210, 255}  % pale blue
\definecolor{constraintgood}{RGB}{255, 210, 80}   % gold

\title[Cosmology with J0946]{Harnessing stellar kinematics to constrain dark energy with the double-source-plane gravitational lens SDSS J0946+1006}

\author[D. J. Ballard et al.]{
Daniel J. Ballard\orcidlink{0009-0003-3198-7151},$^{1}$\thanks{E-mail: daniel.ballard@sydney.edu.au} 
Wolfgang J. R. Enzi\orcidlink{0009-0004-2992-3148},$^{2}$ 
Thomas E. Collett\orcidlink{0000-0001-5564-3140},$^{2}$ 
Coleman M. Krawczyk\orcidlink{0000-0001-9233-2341},$^{2}$ \newauthor 
\ Giovanni Granata\orcidlink{0000-0002-9512-3788},$^{2}$
Tian Li\orcidlink{0009-0005-5008-0381},$^{2}$
Geraint F. Lewis\orcidlink{0000-0003-3081-9319},$^{1}$
Karl Glazebrook\orcidlink{0000-0002-3254-9044},$^{3, 4}$
and Huimin Qu\orcidlink{0009-0006-0299-0265}$^{1}$ %(not final)
\\
$^{1}$Sydney Institute for Astronomy, School of Physics, A28, The University of Sydney, NSW 2006, Australia\\
$^{2}$Institute of Cosmology and Gravitation, University of Portsmouth, Burnaby Rd, Portsmouth PO1 3FX, UK\\
$^{3}$Center for Astrophysics and Computing, Swinburne University of Technology, PO Box 218, Hawthorn, VIC 3122, Australia\\
$^{4}$The ARC Center of Excellence for All Sky Astrophysics in 3 Dimensions (ASTRO 3D), Australia\\
}

\date{Accepted XXX. Received YYY; in original form ZZZ}

\pubyear{\the\year{}}

\begin{document}
\label{firstpage}
\pagerange{\pageref{firstpage}--\pageref{lastpage}}
\maketitle

\begin{abstract}
SDSS J0946+1006 is an attractive target for measuring cosmological parameters through lens modelling. It is the best-studied galaxy-scale strong gravitational lens with multiple sources whose redshift separations are well-suited to constraining the dark energy equation of state. However, multi-plane lens models with free cosmological parameters risk a multi-plane mass-sheet degeneracy, although this can be lifted by a non-lensing deflector density profile tracer. We simultaneously reconstruct near-infrared and near-ultraviolet HST imaging whilst including a velocity dispersion measurement from VLT-MUSE to constrain the foreground deflector. Imaging is reconstructed faithfully regardless of whether the inferred kinematics are realistic, though we find the kinematic constraint essential for shifting the preferred cosmology into a region not in significant tension with other dark energy probes. Deflector density profile perturbations, via substructure and multipolar halo shape deformations, have only a modest effect on inferred cosmology. Combining our fiducial model with Planck CMB data yields $w=-1.01^{+0.08}_{-0.13}$; or, combined with Pantheon SNe Ia, $w=-0.99^{+0.13}_{-0.15}$. We further show that this system proves a remarkably valuable complementary probe in the $w_{0}$-$w_{a}$ plane of an evolving dark energy model, and that the DESI BAO tension with $\Lambda$CDM seen when combined with other datasets is not reproduced when combined with this lens, yielding $(w_{0}, w_{a})=(-0.87^{+0.10}_{-0.11}, -0.22^{+0.28}_{-0.25})$. The system's \textit{third} source, visible with MUSE, only weakly constrains $w$CDM but may strengthen $w_{0}w_{a}$CDM constraints, though further mass-model complexity along its line of sight is required. Overall, kinematics-informed multi-plane lens modelling is a robust route to competitive dark energy constraints, even with a single system.

\end{abstract}


%%%%%%%%%%%%%%%%%%%%%%%%%%%%%%%%%%%%%%%%%%%%%%%

%%%%%%%%%%%%%%%%%%%%%%%%%%%%%%%%%%%%%%%%%%%%%%%%%%

%%%%%%%%%%%%%%%%% BODY OF PAPER %%%%%%%%%%%%%%%%%%

\section{Introduction}
\label{sec:introduction}

\begin{figure*}
    \centering
    \includegraphics[width=\linewidth]{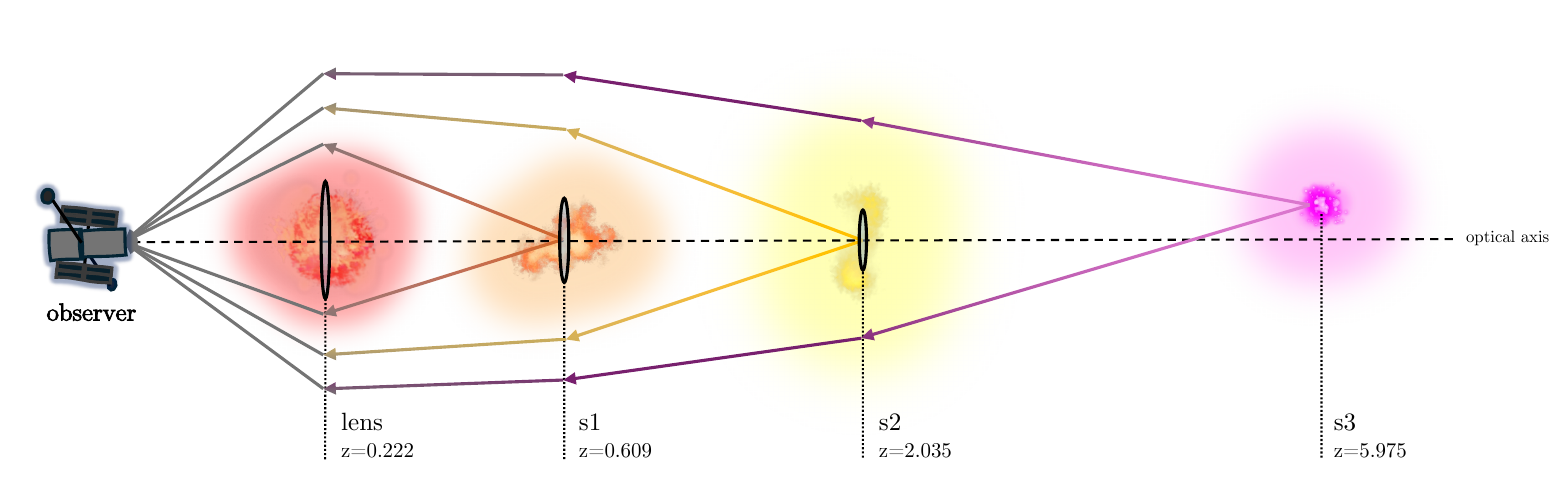}
    \caption{A ray tracing diagram of all four planes along the optical axis in SDSSJ0946+1006. The foreground main deflector (``lens'') deflects the light from s1, s2 and s3; s1 deflects the light from s2 and s3; s2 deflects the light from s3.}
    \label{fig:ray_trace_diagram}
\end{figure*}

Standardisable supernovae at cosmological distances have informed us that the Universe is expanding at an accelerated rate, since they are fainter than anticipated given their distance from us \citep{riess_observational_1998, perlmutter_measurements_1999}. This is physically plausible, provided that $\sim70\%$ of the Universe is filled with a non--luminous fluid exerting negative pressure, known as dark energy. This fluid is considered a cosmological constant, $\Lambda$, which has come to dominate the present-day energy budget of the Universe as all other, non-constant components have diluted with scale factor. Together with the second-most dominant component, Cold Dark Matter (CDM), they form the $\Lambda$CDM cosmological model of our universe.

$\Lambda$CDM is strongly supported by multiple independent probes: Type-1a supernovae \citep[SNe Ia; ][]{brout_pantheon_2022}, as well as galaxy clustering and weak lensing \citep{abbott_dark_2022}, the Cosmic Microwave Background \citep[CMB; ][]{planckcollaboration_planck_2020}, and Baryon Acoustic Oscillations \citep[BAO; ][]{bosscollaboration_cosmological_2015, alam_clustering_2017}. However, recent Dark Energy Spectroscopic Instrument \citep[DESI; ][]{collaboration_overview_2022} BAO results appear to favour a dark energy equation of state that has evolved with redshift \citep{desicollaboration_desi_2025a, desicollaboration_desi_2025}. Since there also exists a persistent tension between early- and late-Universe measurements of the Hubble constant, $H_{0}$ \citep{riess_comprehensive_2022, scolnic_cats_2023, pesce_megamaser_2020, wong_h0licow_2020}, independent tests of $\Lambda$CDM are currently highly desirable.

Other probes of dark energy include strong gravitational lenses, where the curvature of spacetime around foreground masses (deflectors) causes background objects (sources) to appear significantly distorted, magnified, and multiply-imaged. If deflector and source redshifts are known, separations between observed images of sources are determined by the deflector mass distribution and the line-of-sight expansion history. Hierarchical lens modelling of a statistically significant sample of Einstein rings, in combination with deflector stellar kinematics as non-lensing density-profile tracers, can constrain the equation of state parameter for dark energy, $w$, to competitive precision. \citet{li_cosmology_2024} find $w=-0.96\pm0.46$ from 161 lenses, which is compatible with the $w=-1$ scenario, or equivalently a cosmological constant dark energy. They forecast $\sim10\%$ constraints on $w$ from $\mathcal{O}(10^{4})$ lenses observed with Euclid and 4MOST.

Compound lenses, which have $>2$ galaxies aligned along the line of sight, yield observations of multiple Einstein rings through the same foreground density profile, however, thereby more easily disentangling lensing mass from cosmological information. The most ideal such lenses for cosmography are deflectors with both low- and high-redshift sources behind them, which results in significantly different Einstein radii \citep{collett_constraining_2012}. Cluster- and group-scale lenses offer larger cross sections with greater chances to provide this, but have complex mass profiles. It may be possible to obtain meaningful constraints on $w$ on rare occasions when the mass profiles of these systems are well-understood \citep{urcelay_carousel_2026}, but galaxy-scale Double-Source-Plane Lenses (DSPLs), if they can be found, are the most ideal due to typically being the easiest class of strong lenses to model. The Q1 quick data release from Euclid unveiled four such DSPLs \citep{euclidcollaboration_euclid_2025c}, and the AGEL collaboration \citep{tranAGELSurveyStrong2023} discovered six more, although the latter have less favourable redshift configurations for cosmology \citep{barone_agel_2025}. Other notable independent DSPL discoveries include HSC J142449-005322 \citep[“Eye of Horus”;][]{tanaka_spectroscopically_2016}, J1148+1930 \citep[“Cosmic Horseshoe”;][]{belokurov_cosmic_2007}, DES J0408-5354 \citep{shajib_strides_2020}, J0920+4521 \citep{lemon_gravitationally_2023}, and J1721+8842 \citep[the first “Einstein zig-zag”;][]{dux_j1721+8842_2025}. The focus of this work is SDSS J0946+1006 \citep[the “Jackpot Lens”, henceforth J0946;][]{gavazzi_sloan_2008}.

J0946 was spectroscopically selected from the Sloan Digital Sky Survey (SDSS) for HST Advanced Camera for Surveys (ACS) follow-up imaging as part of the Sloan Lens ACS \citep[SLACS; ][]{bolton_sloan_2006, gavazzi_sloan_2008} program. It is described schematically in Figure \ref{fig:ray_trace_diagram}. The $z_{\rm lens}=0.222$ galaxy quadruply-images two spectroscopically-confirmed sources at $z_{\rm s1}=0.609$ and $z_{\rm s2}=2.035$ \citep{smith_fullyspectroscopic_2021}. \citet{collett_cosmological_2014} first used J0946 for DSPL cosmography, finding $w=-1.04\pm0.20$  when combined with CMB constraints and recalibrated using the measurement of $z_{\rm s2}$ in \citet{smith_fullyspectroscopic_2021}\footnote{Originally, \citet{collett_cosmological_2014} found $w=-1.17^{+0.20}_{-0.21}$ by using the older photometric measurement of $z_{\rm s2}=2.41$ from \citet{sonnenfeld_evidence_2012}.}. A third source at $z_{\rm s3}=5.975$, since revealed by the Multi-Unit Spectroscopic Explorer (MUSE) on the Very Large Telescope (VLT), has made J0946 the first galaxy-scale lens with multiply-imaged sources on \textit{three} redshift planes \citep{collett_triple_2020}. This furthest source has not yet been utilised for cosmology.

J0946 also hosts a lensing signature consistent with a $\sim10^9 M_\odot$ dark subhalo \citep{vegetti_detection_2010}, with an anomalously steep central profile \citep{minor_unexpected_2021, nightingale_scanning_2022, ballard_gravitational_2024, despali_detecting_2024, minor_high_2025, tajalli_sharp_2025}, confirmed by joint reconstructions of s1 and s2 \citep{ballard_gravitational_2024, enzi_overconcentrated_2024, minor_high_2025}; \citet{ballard_gravitational_2024} further used s3 image centroids to show milder tension with dark subhalo simulations, while \citet{he_not_2025} largely resolve the tension by allowing degrees of freedom for substructure luminosity. Independent multipole analyses of the main halo shape \citep{minor_unexpected_2021, ballard_gravitational_2024} find evidence for $3^{\rm rd}$- and $4^{\rm th}$-order terms; \citet{enzi_overconcentrated_2024} additionally incorporate the $1^{\rm st}$-order ``lopsidedness'' multipole term \citep{amvrosiadis_lopsidedness_2025}. Altogether, there is strong evidence for mass profiles in J0946 more complex than a simple power law. This complexity has not yet been exploited for dark-energy constraints.

With cosmological parameters held fixed, multi-source-plane models of J0946 yield background sources of near-identical size and shape across different mass models, unlike single-source-plane reconstructions of the same system \citep{ballard_gravitational_2024}, and appear not to significantly invoke a source-position transform \citep{schneider_sourceposition_2014}. However, varying cosmological parameters as extra degrees of freedom is observationally equivalent to a multi-plane mass-sheet degeneracy, as summarised in \citet{teodori_approximate_2026}. While \citet{schneider_can_2014} previously showed this only mildly affects a power-law deflector model of J0946, deviations from a power law could reintroduce the degeneracy, motivating lensing-independent mass tracers. \citet{li_salpeter_2026}, for instance, importance-weight a stellar-plus-dark-matter model using integral-field velocity dispersion data. This only minutely alters the inferred halo slope, though cosmological parameters are not included as degrees of freedom.

Additionally, existing J0946 cosmography constraints in \citet{collett_cosmological_2014} enforced a Uniform prior $w\sim\mathcal{U}(-2, 0)$ on the dark energy equation of state parameter. However, modern analyses from other probes such as \citet{desicollaboration_desi_2025} allow for a much wider prior volume with $w\sim\mathcal{U}(-3, 1)$ in $w$CDM, or $(w_{0}, w_{a})\sim(\mathcal{U}(-3, 1), \mathcal{U}(-3, 2))$ in $w_{0}w_{a}$CDM. Since uncertainties from mass model choices may propagate into the inferred cosmological parameters, adopting a wider, DESI-like prior may better absorb this additional source of uncertainty.

In this paper, we obtain the first constraints on $w$ from to jointly take advantage of multi-plane lensing and kinematics information, using J0946. We model s1 and s2 in two HST bands simultaneously under wider, DESI-like priors for constant and evolving dark energy. We test perturbations to the main deflector's mass model, via substructure and multipolar shape deformations, and investigate the possible constraining power of the third source plane. Section \ref{sec:observations} presents the data; Section \ref{sec:methodology} our methodology; Section \ref{sec:results_lensing_only} constant dark energy results with and without kinematics; Section \ref{sec:w0wa} evolving dark energy. Section \ref{sec:conclusions} concludes.

\section{Observations}
\label{sec:observations}
\begin{figure}
    \centering
    \includegraphics[width=\linewidth]{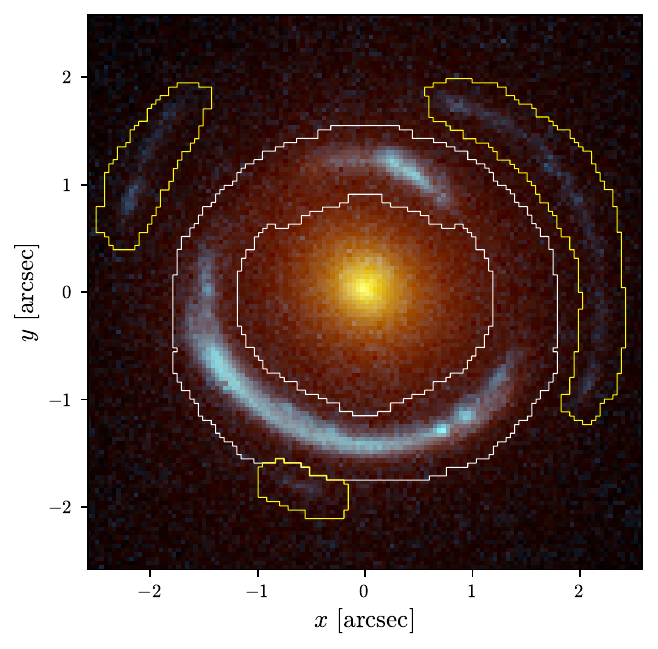}
    \caption{A composite colour image of J0946 using the near-infrared F814W and near-ultraviolet F336W HST imaging of J0946 modelled in this paper. The white and yellow contour shows the pixels masked into the model for reconstructing $s1$ and $s2$, respectively.}
    \label{fig:HST_data}
\end{figure}

\subsection{HST imaging data}
We reconstruct $s1$ and $s2$ from the concentric Einstein rings as they appear in HST data. Specifically, we model an F814W filtered 2096s ACS exposure, drizzled to $0$\arcsec$.05$/pixel, and an F336W filtered 5772s WFC3 exposure, drizzled to $0$\arcsec$.04$/pixel, simultaneously. The F814W data is the same as used in \citet{enzi_overconcentrated_2024}, rather than in e.g. \citet{collett_cosmological_2014, ballard_gravitational_2024}, where the light profile of the main deflector has been pre-subtracted.

We manually paint a pair of non--overlapping masks, which each correspond to pixels used to reconstruct the inner ring or outer ring arcs. Our source reconstructions have no constraining power outside of the pixels contained within these masks. The two-band HST observations are shown as a colour-composite image with masks overlaid in Figure \ref{fig:HST_data}. Our HST PSF is sampled from a star within the field of view of the HST observation, as in \citet{collett_cosmological_2014}.

To account for possible misalignment between the F814W and F336W bands, we include nuisance parameters for a rotation and a translational shift, applied to all mass profile parameters when ray-tracing the second (arbitrarily, F336W) band. Across all models, the resulting coordinate shift between bands is consistently $<1$ pixel.

\subsection{MUSE stellar kinematics of the main foreground deflector}

We utilise the kinematics of the main foreground deflector as measured from integral field spectroscopy data. Specifically, we use a 5-hour integration from MUSE. Following the method outlined in \citet{granata_velocity_2026}, we measure the line-of-sight velocity dispersion of this object, $\sigma_{\rm los, lens}$, measured within a $0$\arcsec$.6$ circular aperture, using the penalized pixel-fitting code \texttt{pPXF} \citep{cappellari_parametric_2004, cappellari_full_2023} and a curated set of X-shooter Spectral Library templates, as in \citet{knabel_tdcosmo_2025}. This yields a measurement of $\sigma_{\rm los, lens}=249.1\pm2.1$ km s$^{-1}$. This is the same measurement utilised in \citet{li_salpeter_2026}.

\section{Methodology}
\label{sec:methodology}
We determine cosmological parameters $\Omega_{m}$ and $w$ to test $w$CDM cosmology, and $\Omega_{m}$, $w_{0}$ and $w_{a}$ to test $w_{0}w_{a}$CDM cosmology, by employing a Bayesian lens model inference pipeline. Cosmological parameters are sampled alongside non--linear parameters for each redshift plane's mass profile and surface brightness distribution, with a likelihood to enforce that the HST imaging data and MUSE kinematics are faithfully reconstructed simultaneously. We forward model the system with \texttt{herculens}\footnote{\href{https://github.com/Herculens/herculens}{https://github.com/Herculens/herculens}} \citep{galan_herculens_2022}, using Stochastic Variational Inference (SVI) followed by Hamiltonian Monte Carlo (HMC), on an Nvidia H100 GPU.

\subsection{Ray tracing}

The lens equation,
\begin{equation}
    \label{eq:lens_equation}
    \boldsymbol{\theta}_{\text{src}}=\boldsymbol{\theta}_{\text{img}} - \boldsymbol{\alpha}(\boldsymbol{\theta}_{\text{img}}),
\end{equation}
states that a 2D image position, $\boldsymbol{\theta}_{\text{img}}$, ray--traces to a 2D position in the source plane, $\boldsymbol{\theta}_{\text{src}}$, by being displaced by a 2D deflection angle, $\boldsymbol{\alpha}(\boldsymbol{\theta}_{\text{img}})$, on the deflector plane.

The deflection angle implicitly contains information about the redshifts $z_{d}$, $z_{s}$ of the deflector and source. Its gradient is proportional to lensing convergence, $\kappa(\boldsymbol\theta)=\frac{1}{2}\boldsymbol{\nabla}\boldsymbol{\alpha}(\boldsymbol\theta)$. This is interpreted as the 2D surface mass density, $\boldsymbol\Sigma(\boldsymbol\theta)$, in units of the 2D surface mass density required for a deflector to multiply-image a source, $\Sigma_{\text{crit}}$, and carries the dependency on the redshifts, $z_{d}$ and $z_{s}$, of the deflector and source:
\begin{equation}
    \label{eq:kappa_in_units_of_sigma_crit}
    \boldsymbol{\kappa}(\boldsymbol\theta; z_{d}, z_{s})=\frac{\boldsymbol{\Sigma}(\boldsymbol\theta)}{\Sigma_{\text{crit}}(z_{d}, z_{s})},
\end{equation}
where
\begin{equation}
\label{eq:sigma_crit}
    \Sigma_{\text{crit}}(z_{d}, z_{s})=\frac{c^{2}}{4\pi G}\frac{d_{A}(z_{\text{s}})}{d_{A}(z_{\text{d}})d_{A}(z_{\text{d}}, z_{\text{s}})},
\end{equation}
and $d_{A}(z_{i}, z_{j})$ denotes the angular diameter distance between two objects $i$ and $j$. The shorthand $d_{A}(z_{j})\equiv d_{A}(0, z_{j})$ denotes the angular diameter distance to an object from the observer.

The intrinsic surface mass density, $\boldsymbol\Sigma(\boldsymbol\theta)$, of a deflector is ignorant to the redshift of the source light being deflected. By assuming that a deflector exists on plane $i$, we consider the $\boldsymbol\Sigma$ observed when the source exists on a plane immediately behind the deflector, index $i+1$, versus when it exists on an arbitrary further plane, index $j > i+1$, and insist that
\begin{equation}
    \boldsymbol{\Sigma}(\boldsymbol{\theta}_{i}; z_{s}=z_{i+1}) = \boldsymbol{\Sigma}(\boldsymbol{\theta}_{i}; z_{s}=z_{j}),
\end{equation}
or, equivalently,
\begin{equation}
    \Sigma_{\text{crit}}(z_{i}, z_{i+1})\boldsymbol{\kappa}(\boldsymbol{\theta}_{i}; z_{i}, z_{i+1}) = 
    \Sigma_{\text{crit}}(z_{i}, z_{j})\boldsymbol{\kappa}(\boldsymbol{\theta}_{i}; z_{i}, z_{j}).
\end{equation}
The convergence, and consequently the deflection angle, is therefore a factor of $\Sigma_{\text{crit}}(z_{i}, z_{j})/\Sigma_{\text{crit}}(z_{i}, z_{i+1})$ greater when measured using a source at redshift $j$ versus redshift $i+1$. We name this factor $\eta_{ij}$ and substitute equation \ref{eq:sigma_crit} to obtain
\begin{equation}
\label{eq:eta}
    \eta_{ij}=\frac{d_{A}(z_{i+1})d_{A}(z_{i}, z_{j})}{d_{A}(z_{j})d_{A}(z_{i}, z_{i+1})}.
\end{equation}
The recursive lens equation, describing where the trajectory of a ray penetrates each $i^{\rm th}$  plane through a system with $j$ source planes behind the foreground-most deflector, is given by:
\begin{equation}
\label{eq:lens_equation_multi_plane}
    \boldsymbol{\theta}_{j} = \boldsymbol{\theta}_{0}-\sum_{i=0}^{j-1}\eta_{ij}\boldsymbol{\alpha}_{i}(\boldsymbol{\theta}_{i}),
\end{equation}
which reduces to Equation \ref{eq:lens_equation} for the $j=1$ case. Throughout this paper, we will refer to the four planes of J0946, $i=\{0, 1, 2, 3\}$, as the foreground deflector-, s1-, s2- and s3-plane respectively.\footnote{In the double-source-plane analysis of J0946 in \citet{collett_cosmological_2014}, there is only one deflection scaling parameter, and it is assigned a different symbol, $\beta$. This follows a deflection scaling convention in \citet{schneider_gravitational_1992}, which defines deflection angles according to the \textit{furthest} source and down-scales deflections to closer sources.}

This paper is primarily concerned with modelling s1 and s2. We therefore require the $j=2$ case of Equation \ref{eq:lens_equation_multi_plane}, such that there is only one $\eta_{ij}\neq1$, which is $\eta_{02}$. For simplicity, we will shorthand $\eta\equiv\eta_{02}$.

\subsection{\textit{Cosmographic} ray tracing}
Cosmological dependence on the ray tracing formalism described enters through $\eta_{ij}$, since the angular diameter distances $d_{A}(z)$ in Equation \ref{eq:eta} are a function of redshift and expansion history. An angular diameter distance is written as:
\begin{equation}
\label{eq:angular_diameter_distance}
    d_{A}(z_{i}, z_{j})=\frac{c}{H_{0}}\frac{1}{1+z_{j}}\left(\frac{1}{\sqrt{|\Omega_{k}|}}\text{sinn}\left(\sqrt{|\Omega_{k}|}\int_{z_{i}}^{z_{j}}\frac{dz}{E(z)}\right)\right);
\end{equation}
\begin{equation}
    \mathrm{sinn}(x) =
    \begin{cases}
      \sinh(x) & \Omega_{k}>0, \\
      x & \Omega_{k}=0, \\
      \sin(x) & \Omega_{k}<0.    \end{cases}
\end{equation}
Note that as a prefactor, $H_{0}$ dependence is divided out once the ratios of distances are taken in Equation \ref{eq:eta}. The deflection angles therefore scale deterministically by redshifts, the energy density due to cosmic curvature, $\Omega_{k}$, and the dimensionless Hubble parameter, $E(z)\equiv \frac{H(z)}{H_{0}}$, which depends on our choice of cosmological model. For $\Lambda$CDM cosmology: 
\begin{equation}
\label{eq:E(z)_for_LCDM}
    E_{\Lambda\text{CDM}}(z)=\sqrt{\Omega_{m}(1+z)^{3}+\Omega_{k}(1+z)^{2}+\Omega_{\Lambda}},
\end{equation}
where $\Omega_{m}$ and $\Omega_{\Lambda}$ are the normalised energy densities of matter (cold dark, as well as baryonic) and dark energy respectively. The more general $w$CDM cosmology yields:
\begin{equation}
\label{eq:E(z)_for_wCDM}
    E_{w\text{CDM}}(z)=\sqrt{\Omega_{m}(1+z)^{3}+\Omega_{k}(1+z)^{2}+\Omega_{\Lambda}(1+z)^{3(1+w)}},
\end{equation}
which reduces to Equation \ref{eq:E(z)_for_LCDM} when $w=-1$. In the Chevallier, Polarski and Linder \citep[CPL; ][]{chevallier_accelerating_2001, linder_exploring_2003} parameterisation, the dark energy equation of state parameter, $w$, has a dependence on scale factor (or, equivalently, redshift):
\begin{equation}
    w(z)=w_{0}+w_{a}(1-a)\equiv w_{0}+\frac{w_{a}z}{1+z}.
\end{equation}
This model, which we refer to as $w_{0}w_{a}$CDM, necessitates a further expansion of Equation \ref{eq:E(z)_for_LCDM}:
\begin{equation}
\label{eq:E(z)_for_w0waCDM}
    E_{w_{0}w_{a}\text{CDM}}(z)=\sqrt{
    \begin{aligned}
        &\ \Omega_{m}(1+z)^{3}+\Omega_{k}(1+z)^{2}+ \\
        &\ \Omega_{\Lambda}(1+z)^{3(1+w_{0}+w_{a})}e^{-3w_{a}z/(1+z)}
    \end{aligned}
    } ,
\end{equation}
which reduces to Equation \ref{eq:E(z)_for_wCDM} when $w_{0}=w$ and $w_{a}=0$. Throughout this work, we assume that our Universe is flat, with $\Omega_{k}=0$.

\subsection{Analytic mass profiles for lens mass and light}
\subsubsection{Elliptical power law}
We model the mass density of the main foreground deflector as an Elliptical Power Law \citep[EPL;][]{tessoreEllipticalPowerLaw2015}, wherein $\rho(r)= r^{-\gamma}$ (for the spherically symmetric case), giving a lensing convergence
\begin{equation}
\label{eq:elliptical_power_law}
    \kappa_{\text{EPL}}=\frac{3-\gamma}{2}\left(\frac{\vartheta_{E}}{qx^{2}+y^{2}/q}\right),
\end{equation}
where $\vartheta_{E}$ is the Einstein radius, $\gamma$ the logarithmic slope, and $q$ the axis ratio; $(x,y)$ are Cartesian coordinates relative to a center $(x_{c},y_{c})$, rotated by position angle $\varphi$.\footnote{$q$ and $\varphi$ are computed deterministically from sampled eccentricities $(e_{1},e_{2})\equiv\frac{1-q}{1+q}(\cos2\varphi,\sin2\varphi)$, avoiding bimodality when sampling $\varphi$ directly.}

We apply this profile to both the main deflector and s1. The s1 origin coordinate is set to the mean de-lensed position of the four brightest points on the inner ring (in the F814W observation), with a Gaussian prior allowing its centroid to deviate from this location.

\subsubsection{External shear}
We include an external shear component, with strength $\Gamma$ and orientation $\varphi_{\Gamma}$\footnote{Defined analogously to $(e_1,e_2)$, with $\Gamma\leftrightarrow1-q$ and $\varphi_\Gamma\leftrightarrow\varphi$.}, accounting for weak lensing effects along line of sight. Following \citet{johnson_lineofsight_2025}, we apply independent external shear parameters on both the foreground deflector and s1 plane.

\subsubsection{Higher--order multipoles}
\citet{etherington_strong_2023} show that the external shear is often stronger than expected versus weak lensing cosmic shear, and preferentially oriented to the position angle of the mass distribution of the lens. This suggests a lensing signal with an internal rather than external origin. We therefore also model for $1^{\rm st}$-, $3^{\rm rd}$- and $4^{\rm th}$-order multipolar expansions to the EPL convergence profile\footnote{The $2^{\rm nd}$-order term is omitted as it is functionally identical to the ellipticity already accounted for in Eq.~\ref{eq:elliptical_power_law}.}:
\begin{equation}
    \kappa_{\mathcal{M}}=\kappa_{\text{EPL}}\times\sum_{n \in \{1, 3, 4\}} A_{\mathcal{M}_{n}}\cos{(n(\varphi-\varphi_{\mathcal{M}_{n}}))}.
\end{equation}
 The strength of the $n^{\text{th}}$ order multipolar contribution is denoted with the amplitude $A_{\mathcal{M}_{n}}$ and its orientation is $\varphi_{\mathcal{M}_{n}}$. The effect is a deviation from a perfect ellipsoidal shape for the halo.

\subsubsection{Substructure}

Due to extensive Bayesian evidence calculations in \citet{vegetti_detection_2010}, \citet{minor_unexpected_2021}, \citet{nightingale_scanning_2022}, \citet{ballard_gravitational_2024}, \citet{despali_detecting_2024} and \citet{minor_high_2025}, we include a truncated Navarro--Frenk--White \citep[tNFW;][]{navarro_structure_1996} lensing perturber, parameterised by a characteristic density, $\rho_{s}$, and scale radius, $r_{s}$, according to
\begin{equation}
    \rho_{\text{tNFW}}=\frac{\rho_{s}}{(r/r_{s})(1+r/r_{s})^{2}}\frac{r_{t}^{2}}{r^{2}+r_{t}^{2}},
\end{equation}
which goes like $\rho_{r}\sim r^{-1}$ when $r<r_{s}$, and $\rho(r)\sim r^{-3}$ when $r>r_{s}$. The truncation, to account for tidal stripping by the halo of the main foreground deflector, follows \citet{baltz_analytic_2009, oguri_detailed_2011}, and smoothly transitions the density profile to $\rho_{r}\sim r^{-5}$ well beyond its truncation radius, $r_{t}$. We place this object at the main deflector redshift, following \citet{enzi_overconcentrated_2024}. And, since \citet{he_not_2025} show that additionally including a light profile for this subhalo yields a better fit to the data, we attach an elliptical Sérsic brightness profile to the same center coordinate. Its parameters are identical between the I- and U-band model, besides amplitude.

\subsubsection{Multi--Gaussian lens light}

The surface brightness of the main deflector is modelled as a sum of elliptical Gaussians, as in \citet{he_unveiling_2024}, but here we do not enforce concentricity or alignment between the components. Through experimentation, we find 20 elliptical Gaussians in each of the two HST bands satisfactory for this object.

\subsection{Forward-modelled pixelated sources}
Rather than a traditional semi--linear inversion method \citep[see e.g.][]{warren_semilinear_2003, vegetti_bayesian_2009}, we propose an analytic lensing mass and pixelated source plane simultaneously, avoiding a lensing matrix inversion altogether. This is made computationally feasible thanks to \texttt{herculens} being written with the auto-differentiable \texttt{JAX} \citep{jax2018github} library.

Our square grid of source pixel brightnesses are drawn from ``coloured'' noise, which has a covariance matrix given by the inverse Fourier transform of the 2D Mátern power spectrum, $\boldsymbol{S}(\omega)$ \citep[see][]{stein_interpolation_1999}, given by:
\begin{equation}
    \boldsymbol{S}(\omega)=4\pi\sigma^{2}n\left(\frac{2n}{\rho^{2}}\right)^{n}\left(\frac{2n}{\rho^{2}}+\omega^{2}\right)^{-(n+1)},
\end{equation}
where $\omega$ is spatial frequency. The physical scale of the source is set by $\sigma^{2}$, its ``smoothness'' is controlled by $n$, and the correlation length is determined by $\rho$. We fit for independent $\sigma$, $n$ and $\rho$ parameters for each source in each band. The real space source is
\begin{equation}
    \boldsymbol{s}(x, y)=\mathcal{F}^{-1}_{\omega\rightarrow(x, y)}\left\{\mathcal{N}\left(0, \sqrt{\boldsymbol{S}(\omega})\right)\right\},
\end{equation}
rescaled with a softplus transformation to enforce positivity:
\begin{equation}
    \boldsymbol{s}(x, y)\leftarrow\left(\log(1+e^{100\boldsymbol{s}(x, y)}\right)/100 ,
\end{equation} 
such that there are no negative pixel brightnesses included within the region of the source planes traced to by our image plane masks.

For s1 we additionally include an elliptical Sérsic component, with its parameters shared between both bands besides amplitude. We find that this efficiently produces a more faithful reconstruction of the inner Einstein ring.

\subsection{Lens model likelihood}

Reconstructing our data with a Bayesian posterior sampling procedure requires us to write a likelihood function, $\mathcal{L}$, for our lens models. Using the analytic profiles described to compute deflection angles and surface brightness distributions, we ray-trace with Equation \ref{eq:lens_equation_multi_plane} and construct a model image, $\boldsymbol{m}$, of dataset $i$, $\boldsymbol{d_{i}}$. We then compare the difference in units of the observational noise of that data, $\boldsymbol\sigma_{\text{obs}, i}$. This comparison, performed jointly on both bands of HST observations in a single, simultaneous fit, forms our likelihood:
\begin{equation}
    \label{eq:likelihood_lensing}
    \mathcal{L}_{\rm lensing}=\prod_{i=\text{F814W}, \text{F336W}}{\mathcal{G}\left(\boldsymbol d_{i}-\boldsymbol m_{i}\left(\xi_{\rm lens}, \xi_{\rm s1}, \xi_{\rm s2}, \xi_{\rm cosmology}\right), \boldsymbol\sigma_{\text{obs}, i}\right)}.
\end{equation}
Our mass and light parameters for the main foreground deflector and s1, and our light parameters for s2, are denoted by $\xi_{\rm lens}$, $\xi_{\rm s1}$ and $\xi_{\rm s2}$ respectively, and our cosmological parameters are denoted by $\xi_{\rm cosmology}$.

Here $\xi_{\rm lens}$, $\xi_{\rm s1}$, $\xi_{\rm s2}$ are the foreground deflector-, s1- and s2-plane parameters, and $\xi_{\rm cosmology}$ the cosmological parameters. We test three foreground deflector mass models: \texttt{EPLsh} (our fiducial model; EPL and external shear), \texttt{EPLshsub} (which additionally includes a tNFW substructure with Sérsic light), and \texttt{EPLshsubmult} (which additionally includes multipoles also). In all cases, $\xi_{\rm lens}$ includes the 20-Gaussian light model; $\xi_{\rm s1}$ is an EPL with external shear with a Sérsic+pixelated light model; $\xi_{\rm s2}$ is pixelated light only.

\subsection{Kinematics likelihood}
\label{sec:kinematics}
In an effort to suppress the multi-plane mass-sheet degeneracy in \citet{teodori_approximate_2026}, we supplement our likelihood using constraints on the stellar velocity dispersion of the main foreground deflector using MUSE. Since this depends on the full 3D deflector potential via the Jeans equations, it is independent of, and complementary to, the projected lensing mass. It can therefore rule out profiles that fit the lensing data via an internal mass-sheet-like rescaling. It cannot, however, constrain uniform convergence \textit{external} to the deflector, which has no dynamical signature.

We compute $\sigma_{\rm los,lens}$ for a given $\vartheta_{E,\rm lens}$, $\gamma_{\rm lens}$ and cosmology via Jeans Anisotropic Modelling (spherical projected \texttt{JAMpy}\footnote{\href{https://pypi.org/project/jampy/}{https://pypi.org/project/jampy/}}; \citealt{cappellari_efficient_2020}). We approximate the EPL profile with 30 concentric Gaussian convergence components; this is then converted to physical surface density using $\Sigma_{\rm crit}$. Because the mass profile is normalised directly from the lensing solution, rather than scaled from the observed light profile, no explicit mass-to-light ratio is assumed. The deflector's surface brightness is already decomposed into Gaussian components in our model by design, providing the luminosity density used as the kinematic tracer in the Jeans equations. The JAM model then integrates the projected second velocity moment of the deflector within a $0$\arcsec$.6$ aperture matching the MUSE aperture in \citet{granata_velocity_2026}, giving $\sigma_{\rm los, lens}\equiv\sqrt{\langle v_{\rm los, lens}^2\rangle}$.

We take this model-predicted velocity dispersion and compare it to the MUSE measurement and its uncertainty, writing our kinematics likelihood as
\begin{equation}
    \label{eq:logL_kin}
    \mathcal{L}_{\rm kinematics}=\mathcal{G}\left(249.1\text{ km s}^{-1}-\sigma_{\rm los, lens}, 2.1\text{ km s}^{-1}\right).
\end{equation}
With the likelihood $\mathcal{L}=\mathcal{L}_{\rm lensing}\mathcal{L}_{\rm kinematics}$ instead of $\mathcal{L}=\mathcal{L}_{\rm lensing}$ only, our three mass models are denoted $\texttt{EPLsh}+\sigma$, $\texttt{EPLshsub}+\sigma$, and $\texttt{EPLshsubmult}+\sigma$.

\subsection{Bayesian inference pipeline}

\subsubsection{Stochastic Variational Inference}
We first implement the \texttt{numpyro} \citep{phan_composable_2019} Stochastic Variational Inference (SVI) algorithm to fit a multivariate Gaussian--shaped posterior to our model parameters. This processes minimises the Kullback--Leibler (KL) divergence, to match the true posterior to this more tractable one as closely as possible \citep{blei_variational_2017}, using the \texttt{AdaBelief} optimiser \citep{zhuang_adabelief_2020} to do so. At this stage, we implement only the lensing likelihood for all model cases. We run this SVI procedure 30 times for 50,000 iterations each for every model fit performed. 

\subsubsection{Hamiltonian Monte Carlo}

We initialise a Hamiltonian Monte Carlo (HMC) sampler from the median sample of the lowest-KL SVI run, using the \texttt{numpyro} No U-Turns Sampler (NUTS) within a Gibbs routine \citep{krawczyk_ckrawczyk_2024} that samples the light and the mass (which also includes the cosmological) parameters independently. Four chains run simultaneously until $\hat{r}<1.05$; for computational cost, the multi-Gaussian deflector surface brightness profiles in both bands are held fixed at their SVI values. Since \texttt{JAMpy} is not \texttt{JAX}-compatible, we pre-compute a look-up table of $\sigma_{\rm model}$ over our priors on $\vartheta_E^{\rm lens}$, $\gamma^{\rm lens}$, $\Omega_m$ and $w$ (or $w_0$ and $w_a$), given the SVI surface brightness distribution of the foreground deflector, and interpolate over this table during HMC to evaluate the kinematics likelihood.

\subsection{Testing our methodology}
\label{sec:mock}

A similar pipeline was used on this lens in \citet{enzi_overconcentrated_2024}. As this is its first use for cosmological constraints, however, we verify that it can recover an $\eta$ value as agreeable with $\Lambda$CDM as found by independent, semi-linear inversion code in \citet{collett_cosmological_2014}. We first model the lens-light-subtracted F814W data used in \citet{collett_cosmological_2014}, with similar assumptions: an EPL with external shear on the foreground deflector, and an SIS with fixed centroid at the mean s1-plane position of four conjugate points on the inner arcs. From the chain sample with $\eta$ closest to their median $\beta^{-1}$, we generate synthetic data by applying J0946's observed noise map, modulated by Gaussian noise $\sim\mathcal{N}(0.0, 0.1)$, to the predicted arcs, then re-model this synthetic data under the same assumptions and priors. This recovers all observable features to noise level, with $\eta$ recovered within $1\sigma$ of the resulting posterior.

\begin{table*}
    \centering
    \begin{tabular}{|l|cccc|}
    \hline
    Model & $\left\langle \chi^{2} \right\rangle$ 
      & \makecell{$\Delta\text{WAIC}$}
      & \makecell{$\sigma_{\rm los, lens}$ [km/s]\\(tension [$N\sigma$])}
      & \makecell{$\eta$ tension [$N\sigma$]\\CMB / SNe Ia} \\
    \hhline{|=|====|}
    \texttt{EPLsh}                 & 1.07 & $136.4 \pm 43.7$ %1.025
    & $264.1^{+7.7}_{-6.3}$               (2.3) & 3.2 / 2.3          \\
    \texttt{EPLshsub}              & 1.04 & $165.2 \pm 69.3$ %0.995
    & $265.8^{+5.0}_{-4.8}$               (3.2) & 3.6 / 3.5          \\
    \texttt{EPLshsubmult}          & 1.05 & $383.5 \pm 70.1$ %1.003
    & $284.0^{+13.6}_{-11.5}$             (3.0) & 3.6 / 3.4          \\ \hline
    \texttt{EPLsh}$+\sigma$        & 1.08 & --- %1.027
    & $250.5\pm1.9$                       (0.5) & 2.3 / 0.6          \\
    \texttt{EPLshsub}$+\sigma$     & 1.05 & $255.7 \pm 73.3$ %1.010
    & $252.3\pm1.9$                       (1.1) & 3.0 / 2.2          \\
    \texttt{EPLshsubmult}$+\sigma$ & 1.05 & $299.3 \pm 53.8$ %1.002
    & $251.1\pm2.0$                       (0.7) & 2.5 / 1.1          \\ \hline 
\end{tabular}
    \caption{Summary statistics for all lens models, with and without a kinematics constraint. Columns show: the average $\chi^{2}$ statistic between the data and the highest-likelihood reconstruction, computed over all pixels within the s1 and s2 masks; the significance of the model's imaging predictive accuracy relative to our fiducial \texttt{EPLsh}$+\sigma$, case, $\Delta\mathrm{WAIC}$; the model-predicted line-of-sight velocity dispersion of the foreground deflector, $\sigma_{\rm los,lens}$, and the $N\sigma$ tension between the predicted and measured velocity dispersion of the foreground deflector; the $N\sigma$ tension in $\eta$ versus the expectation from Planck CMB and Pantheon SNe Ia.}
    \label{tab:model_comparison}
\end{table*}
\section{Constraints on constant dark energy}
\label{sec:results}
\label{sec:results_lensing_only}
\label{sec:results_lensing_kinematics}
\begin{figure}
    \centering
    \includegraphics[width=\linewidth]{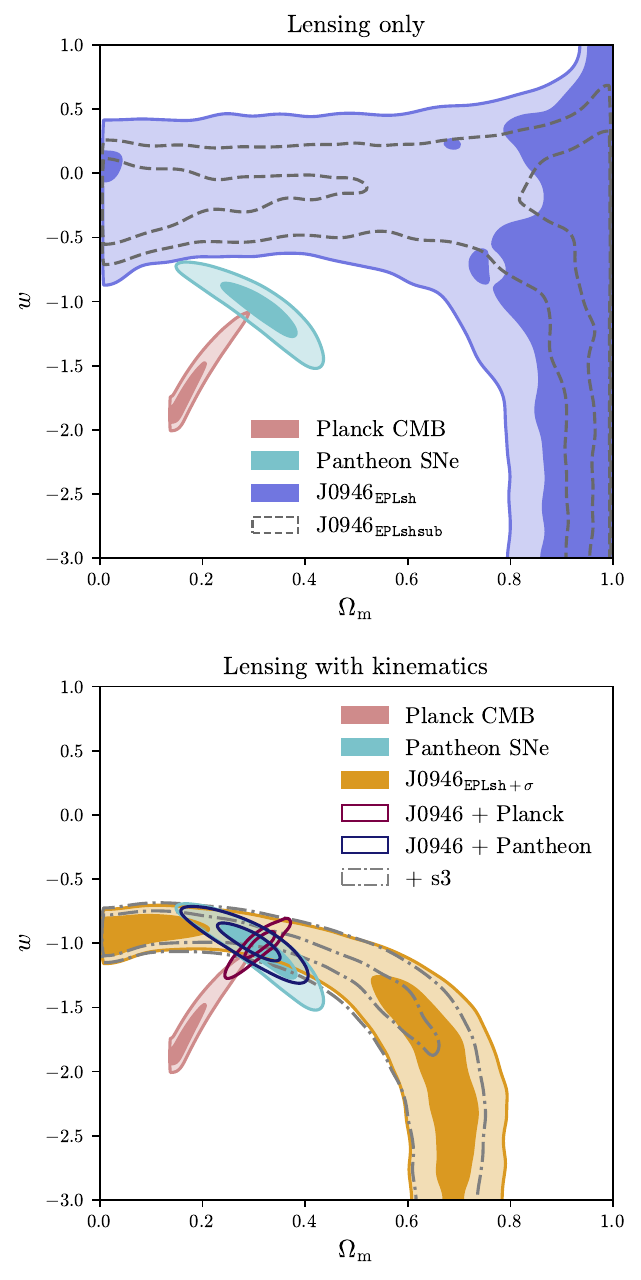}
    \caption{Posteriors on $\Omega_{m}$ and $w$ for our fiducial \texttt{EPLsh} model, without (upper, blue) and with (lower, orange) stellar kinematics, alongside Planck CMB \citep[pink;][]{planckcollaboration_planck_2020} and Pantheon SNe Ia \citep[cyan;][]{scolnic_complete_2018} constraints. The lower panel also shows joint posteriors with J0946$_{\texttt{EPLsh}+\sigma}$ (purple, indigo). Dashed and dash-dotted contours show the effect of adding substructure and re-weighting by an s3 focusing term, respectively. Contours denote $1\sigma$ and $2\sigma$ confidence regions.}
    \label{fig:wCDM_with_without_kinematics}
\end{figure}
Here we present and discuss our $w$CDM constraints from J0946, modelled with the lensing likelihood alone and jointly with the kinematic likelihood. Our main results are summarised in Figure \ref{fig:wCDM_with_without_kinematics}, which shows $\Omega_{m}$-$w$ posteriors for J0946 with and without kinematics, overlaid with Planck CMB \citep{planckcollaboration_planck_2020} and Pantheon SNe Ia \citep{scolnic_complete_2018} constraints. Individual lens model inferences are briefly discussed are given in Appendix \ref{app:individual_lens_models}.

\subsection{Constraining $w$CDM with lensing only}
By constraining cosmology alongside the lens model with the HST imaging data alone, we highlight a tension between our J0946 model and existing constraints from both Planck \citep{planckcollaboration_planck_2020} and Pantheon \citep{scolnic_complete_2018}. We therefore do not compute joint posteriors with these probes, as the overlap between them occurs well into the tails of both datasets. We find that the addition of a substructure to our fiducial J0946$_{\texttt{EPLsh}}$ model does result in a tighter constraint in $\Omega_{m}$-$w$ space, but still in tension. We show the J0946$_{\texttt{EPLshsub}}$ posterior contours in this plane, compared with the fiducial case, in the upper panel of Figure \ref{fig:wCDM_with_without_kinematics}. The addition of multipoles as well does not constrain this region any more tightly than this, however.

We note that this tension was not observed in previous results in this system \citep{collett_cosmological_2014}, though those constraints were less precise, and we required a wider prior on $w$ in order to access the likelihood peak that our model preferred. It is also possible that other differences - such as the extra complexity we afforded to the s1-plane mass distribution, or our inclusion of the F336W data - could be contributing to this disagreement.

\subsection{Constraining $w$CDM with lensing and kinematics}

The inclusion of a kinematic constraint, however, shows that this possible tension between J0946 and other dark energy probes may not be genuine. With non-lensing, density profile-constraining data included, we are restricted to a region far into the tail of our lensing-only posterior in $\Omega_{m}$-$w$ space, as shown in the lower panel of Figure \ref{fig:wCDM_with_without_kinematics}. The far slimmer posterior distribution in this plane, as well as its orientation, makes this a highly effective complementary probe to Pantheon, and particularly Planck. We report a constraint from our fiducial, J0946$_{\texttt{EPLsh}+\sigma}$ model as $\Omega_{m}=0.30\pm0.06$ and $w=-0.99^{+0.13}_{-0.15}$ when combined with Pantheon, or $\Omega_{m}=0.31\pm0.03$ and $w=-1.01^{+0.08}_{-0.13}$ when combined with Planck. These results are comfortably consistent with the concordance $\Lambda$CDM model.

With this in mind, we conclude that kinematics data is essential to extracting cosmological information from (at least \textit{this}) double-source-plane lens. We present all inferred 1D posterior distributions on $w$ and $\Omega_{m}$ in Appendix \ref{sec:ridge_plots_wCDM}, which contains our fiducial kinematics-informed lens model scenario, as well as the substructure and multipole-perturbed cases, alone and combined with Planck CMB and Pantheon SNe Ia, as well as DESI BAO data. We note that our J0946 results add little value to $w$CDM constraints with DESI, which already constrains $w$ with $\sim7\%$ precision by itself.

\subsection{Does mass model choice matter?}
\begin{figure}
    \centering
    \includegraphics[width=\linewidth]{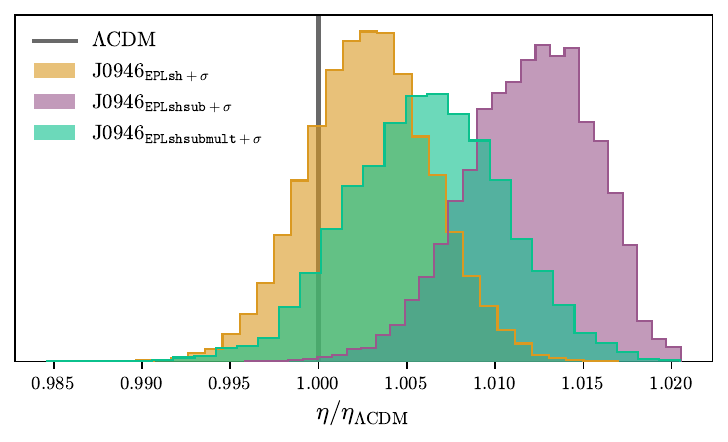}
    \caption{The inferred posterior distribution on $\eta$, normalised by the $\Lambda$CDM-predicted value, for each of our J0946$_{*+\sigma}$ models.}
    \label{fig:eta_comparison}
\end{figure}

\begin{figure*}
    \centering
    \includegraphics[width=\linewidth]{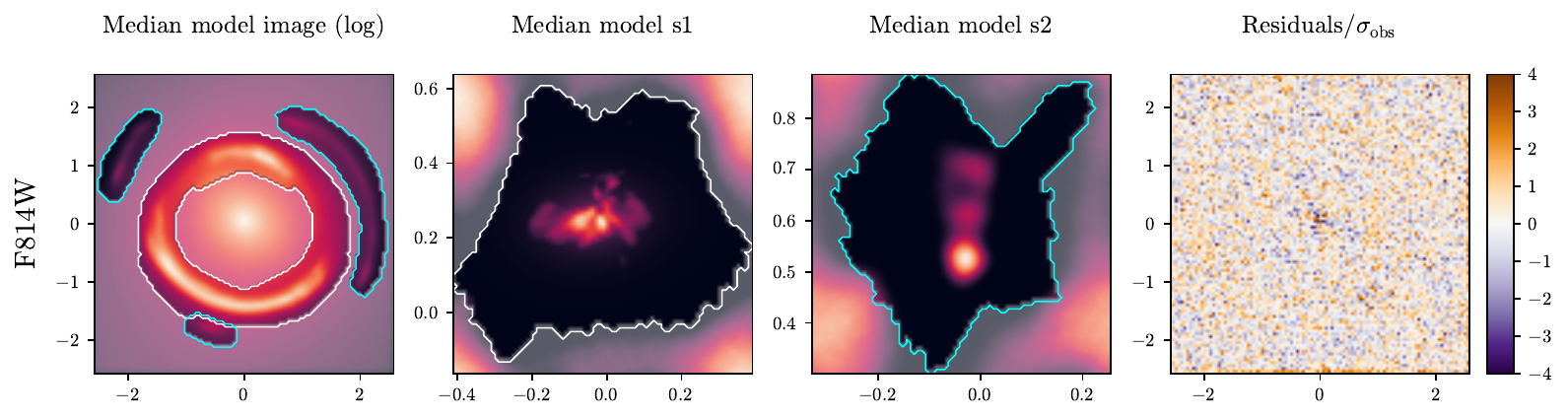}
    \includegraphics[width=\linewidth]{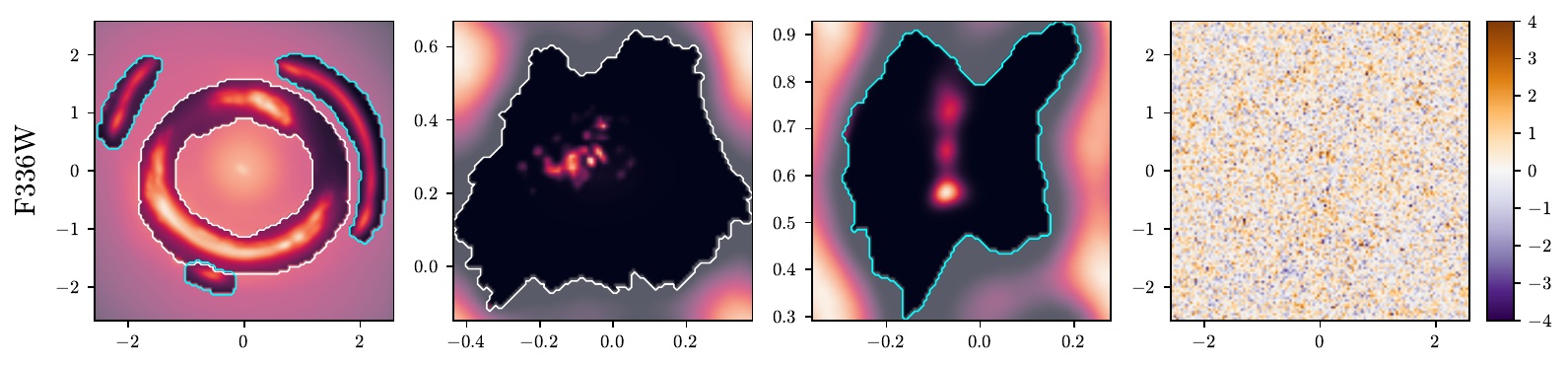}
    \caption{An example reconstruction of the F814W (top) and F336W (bottom) data, selected from the median of our \texttt{EPLshsubmult}$+\sigma$ model. Panels left-to-right show reconstructions at this sample of the image plane, s1 plane, s2 plane, and residuals between the observation and image plane model.}
    \label{fig:model_reconstructions_814}
\end{figure*}
\begin{figure}
    \centering
    \includegraphics[width=\linewidth]{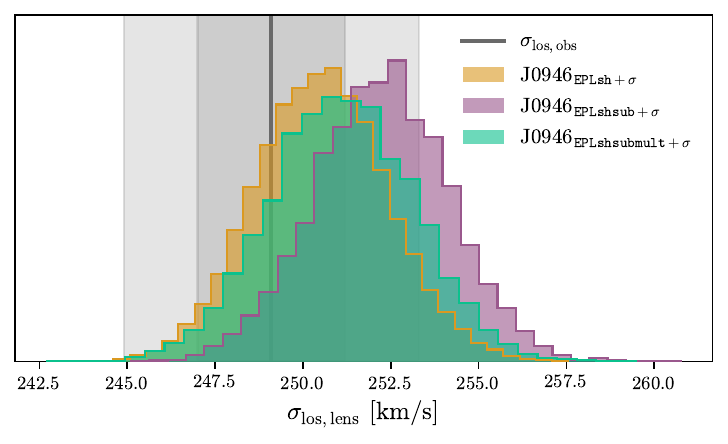}
    \caption{Posterior distributions on $\sigma_{\rm los, lens}$ for the \texttt{EPLsh}$+\sigma$, \texttt{EPLshsub}$+\sigma$ and \texttt{EPLshsubmult}$+\sigma$ models. Vertical grey line and shaded bands denote $\sigma_{\rm los, lens}$ as measured with VLT--MUSE with $1\sigma$ and $2\sigma$ uncertainty.}
    \label{fig:sigma_los_lens_results}
\end{figure}
The inferred value of $\eta$ for our three kinematics-informed models is shown in Figure \ref{fig:eta_comparison}. Out of the mass models we have tested for in this paper, there is a $\sim2\%$ shift, due to which foreground deflector mass model, which remains to be accounted for. We record the $N\sigma$ tensions between our $\eta$ posteriors from J0946 modelling and $\eta$ posteriors expected from $\Omega_{m}$ and $w$ according to Planck and Pantheon in Table \ref{tab:model_comparison}. Across all mass models, including the kinematic likelihood systematically reduces the tension in $\eta$ with both CMB and SNe Ia; the \texttt{EPLshsub}$+\sigma$ case is a slight outlier, remaining in only marginally less tension with existing probes versus our fiducial, lensing likelihood-only case. We note that the parts of these distributions where there is overlap with Planck CMB and Pantheon SNe Ia data in the $\Omega_{m}$-$w$ plane are still consistent enough across the different models that our conclusions about the nature of dark energy from J0946 do not significantly change. The resulting shift in 1D error bars of the cosmological parameters as we combine each model with existing probes can be observed in Appendix \ref{sec:ridge_plots_wCDM}.

The goodness-of-fit between our models is summarised in Table \ref{tab:model_comparison}. For each of our model cases, we report the average $\chi^{2}$ statistic per-pixel, denoted as $\left\langle\chi^{2}\right\rangle$, computed between the data and the highest-$\mathcal{L}$ model image within the s1 and s2 pixel masks, averaged over both bands. For a well-fitting model with accurate noise estimates, the expectation value of $\chi^{2}$ in each pixel is unity, so $\left\langle \chi^{2} \right\rangle \approx 1$ indicates a good fit; given $N_{\rm pixels}=9135$ pixels within our masks, this statistic has a noise floor of $\sqrt{2/N_{\rm pixels}}\approx0.01$. We caution, that the drizzling used to construct our data introduces correlated noise between neighbouring pixels, which this per-pixel statistic does not account for; $\left\langle\chi^{2}\right\rangle$ should therefore be interpreted as a qualitative, relative indicator of fit quality across models, rather than a strictly calibrated absolute one.

With this in mind, all models in Table \ref{tab:model_comparison} satisfy $\left\langle \chi^{2} \right\rangle \approx 1$, indicating that the imaging data alone do not more or less strongly discriminate between mass model choices, whether kinematics are included or not. We consistently produce near-noise level residuals between model and data in all lens models tested. 

Because $\left\langle\chi^{2}\right\rangle$ alone cannot meaningfully distinguish between models at this level of fit quality, we additionally report the difference in the Widely Applicable Information Criterion, \citep[$\Delta\text{WAIC}$;][]{watanabe_asymptotic_2010}, relative to our fiducial model. Unlike $\left\langle\chi^{2}\right\rangle$, which only assesses agreement between the best-fit model and the data, WAIC estimates the expected predictive accuracy of the full posterior on new data, and includes a penalty term that punishes according to the number of degrees of freedom. A larger $\Delta\text{WAIC}$ therefore indicates a model with greater predictive accuracy relative to the fiducial model, providing a more sensitive comparison than $\left\langle\chi^{2}\right\rangle$ where imaging residuals alone are similar across models. By this metric, \texttt{EPLshsubmult} is mildly preferred among the lensing-only models, and \texttt{EPLshsubmult}$+\sigma$ among the kinematics-informed models, though the margin between the two best-performing models in each case does not constitute a statistically confident preference given their respective uncertainties.

Consistently with the $\left\langle\chi^{2}\right\rangle$ statistic, we find with $\Delta\text{WAIC}$ that the addition of substructure and multipoles produces a marginally improved fit over the fiducial model, as in existing works \citep[e.g.][]{ballard_gravitational_2024}, but the model is indistinguishably improved between lensing-only and kinematics-informed reconstructions with the same mass model assumptions.

We also calculate the $\sigma_{\rm los, lens}$ predicted by our lensing-only models, and find that they are in $2-3\sigma$ tension with the measured value from MUSE, for all three model variants. Including the kinematics constraint necessarily brings this tension down to $\lesssim1\sigma$ in all cases. This discrepancy in predicted kinematics between the lensing-only and kinematics-informed models, despite both reproducing an indistinguishable imaging fit, is a tangible observational signature of the mass-sheet degeneracy. We therefore interpret our kinematic models as suppressing this degeneracy, and find them to be more trustworthy reconstructions of the data.

We demonstrate the fidelity of our simultaneously reconstructed HST observations in Figure \ref{fig:model_reconstructions_814}, where we show our models of the F814W and F336W HST data, and predicted source brightness profiles for the median of the lowest-$\chi^{2}$ model, J0946$_{\texttt{EPLshsubmult}+\sigma}$. We also show consistency between $\sigma_{\rm los, lens}$ predicted by our J0946$_{*+\sigma}$ models and the MUSE data in Figure \ref{fig:sigma_los_lens_results}. All three prefer a velocity dispersion $\sim1\sigma$ greater than measured, which is not a statistically significant finding. We do however note that a degeneracy between $\eta$ and $\sigma_{\rm los, lens}$ means that the model in greater tension with other cosmology probes (J0946$_{\texttt{EPLshsub}+\sigma}$) also shows greatest tension with the kinematics data (lensing-only models notwithstanding).

It is noteworthy that our inferred halo slope in all of our models is sub-isothermal ($\gamma<2$) unlike in \citet[e.g.][]{collett_cosmological_2014}. A slightly sub-isothermal slope ($\gamma=1.96\pm0.02$) is also found in \citet{turner_twodimensional_2024}, who combine spatially-resolved kinematics derived from the MUSE data with an imposed lensing mass constraint at the Einstein radius. Broadly, dynamics constrain the 3-dimensional radial mass profile in a way that projected lensing observables alone cannot, and is preferentially sensitive to the mass enclosed within a different characteristic radius, allowing it to probe the central mass distribution more effectively than with lensing. This may help to explain why the inferred slope in our work and in \citet{turner_twodimensional_2024} differs from previous lensing-only models of this object.

We also note that s1 is consistently super-isothermal, such that a trade-off between cuspiness on one plane versus the other may be at play. It is possible that
a coupled mass-sheet-like degeneracy between the two lens planes, as derived by \citet{johnson_degeneracies_2026}, may partly explain this. To our knowledge, all previous works on J0946 have enforced $\gamma=2$ on this object, however, and little is known about the most appropriate mass profile to use for an irregular $z=0.609$ galaxy. A study into mass model complexities on this plane will be the subject of future work.

\subsection{Does the third source plane provide $w$CDM constraining power?}
We also investigate how well the extra source from MUSE, s3, can constrain both $w$CDM and $w_{0}w_{a}$CDM, utilising our existing J0946$_{\texttt{EPLsh}+\sigma}$ models.  We achieve this by fitting the centroid position of the two images of s3, which we refer to as s3$_{\rm A}$ and s3$_{\rm B}$. We treat these two centroids and their uncertainties as distributions from which to draw 10,000 points to ray trace through to the s3-plane at every sample in our J0946$_{\texttt{EPLsh}+\sigma}$ model. We then re-weight our lens model chain according to the region of overlap between the two resulting clouds of points, by fitting a KDE to each cloud and evaluating the s3$_{\rm A}$ KDE and s3$_{\rm B}$
KDE at the s3$_{\rm B}$ point cloud and s3$_{\rm A}$ point cloud, respectively. The average of all of these KDE evaluations at each J0946$_{\texttt{EPLsh}+\sigma}$ sample is taken as that sample's weight, once it has been normalised using all other resultant weights across the whole chain.

 The s3-weighted J0946$_{\texttt{EPLsh}+\sigma}$ posterior is overlaid in the lower panel of Figure \ref{fig:wCDM_with_without_kinematics}; samples with particularly low $w$ and particularly high $\Omega_{m}$ are down-weighted, but this region holds little importance when combining with existing probes which already heavily disfavour these combinations of values. We therefore conclude that the addition of s3 in this way yields very little extra constraining power in $w$CDM.

\section{Constraints on evolving dark energy}
\label{sec:w0wa}
We re-perform our fiducial, J0946$_{\texttt{EPLsh}+\sigma}$ model by substituting Equation \ref{eq:E(z)_for_w0waCDM} into Equation \ref{eq:angular_diameter_distance} in order to calculate $\eta$ with Equation \ref{eq:eta}, thereby incorporating $w_{a}$ as an additional free parameter. Having used the same priors as used in the analyses in \citet{desicollaboration_desi_2025} by design, we pay particularly close attention to the $w_{0}$--$w_{a}$ plane when combined with this probe. The results of J0946$_{\texttt{EPLsh}+\sigma}$ in combination with Pantheon SNe Ia \citep{scolnic_complete_2018} as well as DESI, in the $w_{0}$-$w_{a}$ and $\Omega_{m}$-$w_{0}$ planes are shown in Figure \ref{fig:w0wa_EPLshkin}.

Joint posteriors between J0946 and Pantheon supernovae yields $\Omega_{m}=0.30^{+0.06}_{-0.07}$, $w_{0}=-1.00^{+0.15}_{-0.19}$ and $w_{a}=-0.00^{+0.21}_{-0.29}$, whereas a joint posterior between J0946 and DESI yields $\Omega_{m}=0.30\pm0.01$, $w_{0}=-0.87^{+0.10}_{-0.11}$ and $w_{a}=-0.22^{+0.29}_{-0.25}$. Unlike in $\Omega_{m}$-$w_{0}$ space (or $\Omega_{m}$-$w$ space previously), the area spanned by the J0946$_{\texttt{EPLsh}+\sigma}$ posterior contours in $w_{0}$-$w_{a}$ space is comparable in size to those of Pantheon and DESI \textit{before} combining them, making this a competitive probe even by itself, with $w_{0}=-1.52^{+0.60}_{-0.96}$ and $w_{a}=-0.26^{+0.42}_{-0.28}$. It is particularly noteworthy that whilst Planck and DESI are precise probes of $w_{0}$ but less precise probes of $w_{a}$, the opposite is true for J0946, which measures $w_{0}$ much less precisely but $w_{a}$ much more precisely. This makes this lens a remarkably valuable complementary probe for constraining $w_{0}w_{a}$CDM cosmology. However, it is currently unknown whether this constraining power will be a generic property of DSPLs once a statistically significant sample is obtained; it may be that J0946 is an outlier, with particularly favourable redshifts for constraining the evolution of dark energy. When combining J0946 with DESI, we obtain a joint posterior with $\Omega_{m}=0.31\pm0.01$, $w_{0}=-0.87^{+0.10}_{-0.11}$, $w_{a}=-0.22^{+0.28}_{-0.25}$. Therefore, whereas other dark energy probes combined with the same DESI posterior yield $3-4\sigma$ tensions with $\Lambda$CDM \citep{desicollaboration_desi_2025}, we see a far less statistically significant $\lesssim1\sigma$ tension when combining J0946 with DESI BAO.

\subsection{Does the third source plane better constrain $w_{a}$?}
\label{sec:source_3}
\begin{figure}
    \centering
    \includegraphics[width=0.9\linewidth]{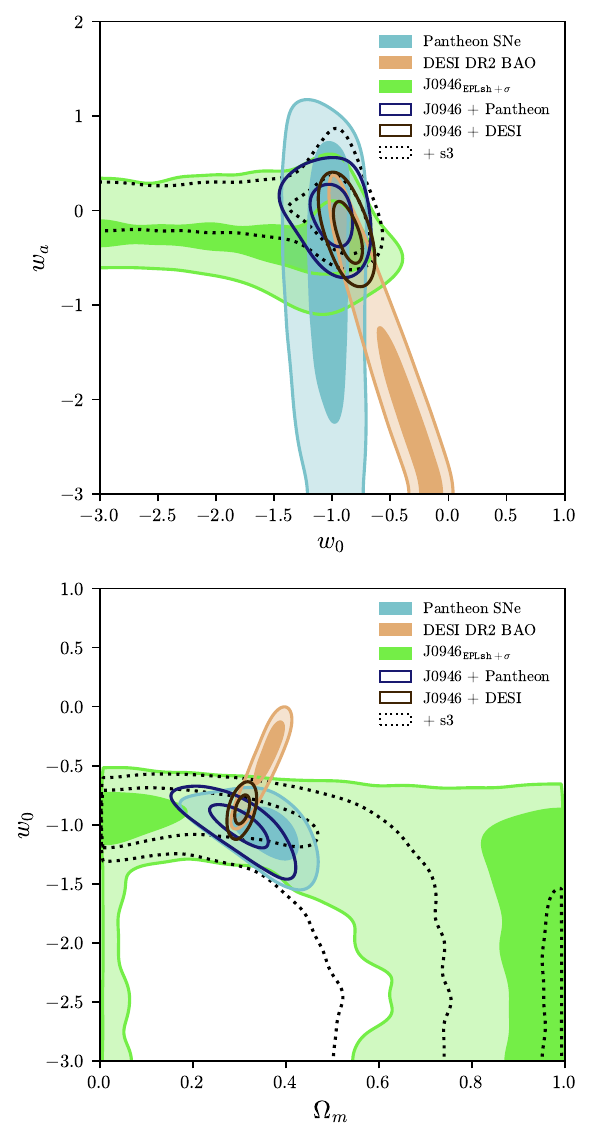}
    \caption{Posterior distributions in the $w_{0}$--$w_{a}$ plane (top) and $\Omega_{m}$--$w_{0}$ plane (bottom) for a J0946$_{\texttt{EPLsh}+\sigma}$ mass model with a $w_{0}w_{a}$CDM prior. Green, turquoise and orange filled contours show J0946$_{\texttt{EPLsh}+\sigma}$, Pantheon SNe Ia and DESI BAO constraints respectively. The unfilled navy and mahogany contours show joint posteriors Pantheon+J0946$_{\texttt{EPLsh}+\sigma}$ and DESI+J0946$_{\texttt{EPLsh}+\sigma}$ respectively. Contours denote $1\sigma$ and $2\sigma$ confidence levels.}
    \label{fig:w0wa_EPLshkin}
\end{figure}
\begin{figure}
    \centering
    \includegraphics[width=0.95\linewidth]{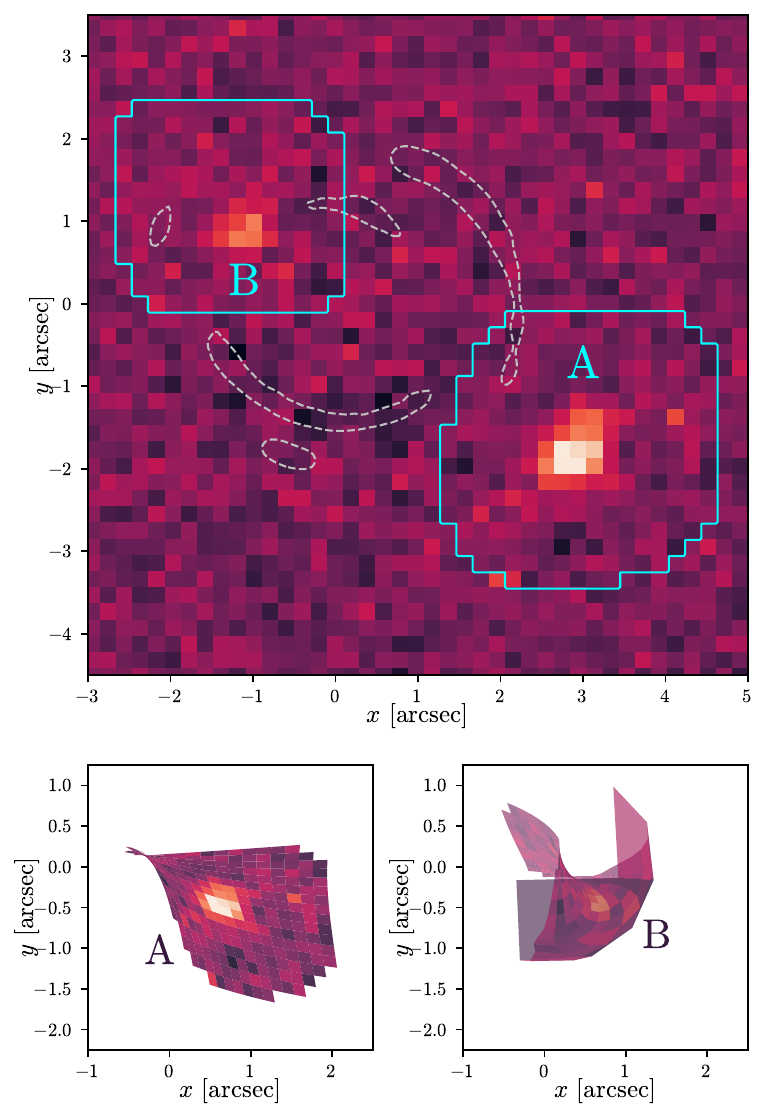}
    \caption{Continuum-subtracted 5\AA\ window around 8475\AA\ of the MUSE data, where the two images of s3 are visible (top panel). Dashed contours represent the location of s1 and s2 in HST relative to this data. Solid cyan contours denote the region of pixels included when obtaining de-lensed projections of images A and B in the s3-plane (bottom panels).}
    \label{fig:s3_delensing}
\end{figure}
 We find s3 to be far more informative for our $w_{0}w_{a}$CDM model than in our $w$CDM model previously. We find that samples that are more agreeable with the concordant $w_{a}=0$ scenario are up-weighted by the focusing of s3, with $w_{0}=-1.11^{+0.24}_{-1.08}$ and $w_{a}=0.00^{+0.20}_{-0.26}$, from J0946$_{\texttt{EPLsh}+\sigma}$ (+s3) alone. This re-weighted posterior is overlaid on our un-weighted model in Figure \ref{fig:w0wa_EPLshkin}. In Appendix \ref{sec:ridge_plots_w0waCDM}, we show our 1D cosmological parameter posterior distributions as combined with Planck CMB, Pantheon SNe Ia and DESI BAO constraints, with and without the s3 focusing weights applied.

We wish to highlight, however, that our methodology for incorporating s3 in this analysis is imperfect. Although it is reasonable to assume \textit{a priori} that the two measured centroids of the images should map to the same source plane location, it should also be true that the brightness distribution of each image, projected back to the s3-plane, should match. In Figure \ref{fig:s3_delensing}, we show that whilst the two de-lensed images are projected to the same position, their predicted source plane morphologies disagree, with the two images each predicting a source in the same position but of a different shape and size. Additionally, for image B, an extra folding of this region of the image plane appears to have occurred, which would imply that image B is further split into two images when ray-traced forwards from the s3-plane, perhaps in a ``zig-zag''-like compound lensing scenario \citep{collett_compound_2016, dux_j1721+8842_2025, salama_microlensing_2026}. We also have not included a mass distribution on s2 in this analysis, though it is noteworthy that s2 and s3 are not closely aligned along one line of sight, and s2 is known to have a negligible lensing effect on s3 \citep{smith_fullyspectroscopic_2021, ballard_gravitational_2024}. It is clear that current lens models of the line of sight between the image plane and s3 are insufficient; this will be explored in a future study.

\section{Conclusions}
\label{sec:conclusions}

Strong gravitational lenses with multiple background sources offer a compelling, independent route to constraining the dark energy equation of state, but multi-plane lens models risk introducing a mass-sheet degeneracy once cosmological parameters are allowed to vary, yielding untrustworthy cosmological constraints. In this work, we set out to establish whether jointly modelling a kinematics constraint of the multi-source-plane SDSS J0946+1006 strong lens system, alongside a reconstruction of the imaging data, can practically deliver dark energy constraints that are more robust and genuinely competitive with existing cosmological probes.

We performed fully Bayesian dark energy cosmography with the first two source planes of J0946, using F336W imaging alongside F814W, and implemented a likelihood term that constrains the predicted velocity dispersion of the lens model using a measurement of $\sigma_{\rm los, lens}=249.1\pm2.1$ km s$^{-1}$ from MUSE integral field spectroscopy. Our cosmological posteriors were found to be in significant tension with $\Lambda$CDM unless stellar kinematics were used as an external tracer of the 3D halo profile. Whilst our models were near-indistinguishably good at reconstructing the HST observations whether kinematics were included as a constraint or not, we found that reasonable lens model priors contain regions with unrealistic foreground deflector velocity dispersions. The combined lensing-plus-kinematics likelihood, however, peaked in a region where the resulting cosmological parameter posteriors provide meaningful complementary information to Planck CMB \citep{planckcollaboration_planck_2020} and Pantheon SNe Ia data \citep{scolnic_complete_2018}, and also gave markedly more precise constraints than DSPL cosmography without kinematics.

We discussed that J0946 has already been extensively modelled to constrain the foreground deflector's mass distribution beyond a simple power-law profile, with several single-, double- and triple-source-plane analyses concluding that dark matter substructure and multipolar halo deformations are required for optimal fits \citep[e.g.][]{minor_unexpected_2021, ballard_gravitational_2024, he_not_2025, tajalli_sharp_2025}. These findings had not previously been used to constrain cosmology in this system, nor to investigate more generally how to perform cosmography beyond the simplistic power-law mass model that is known to suppress the multi-plane mass-sheet degeneracy here \citep{schneider_can_2014}. We found that the posterior shapes in the $\Omega_{m}$-$w$ plane of a $w$CDM cosmological model change only modestly, particularly where they intersect other cosmological probes for joint constraints. We perturbed the power-law halo model of the foreground deflector using a substructure alone, and with multipolar shape deformations also, and observed that the inferred cosmology-tracing $\eta$ parameter could drift by $\sim2\%$. When a substructure was added but multipoles were not, the model was least consistent with the concordant cosmology, and also yielded a less accurate kinematic description of the foreground deflector.

Extending our fiducial, unperturbed lens model to include $w_{a}$ as a free parameter gave posteriors in $w_{0}$--$w_{a}$ space that were near-orthogonal to Pantheon supernovae and DESI BAO. We found that combining J0946 with DESI yielded posteriors $\sim1\sigma$ consistent with $\Lambda$CDM, unlike when DESI has been combined with other probes \citep{desicollaboration_desi_2025}.

A rudimentary \textit{triple}-source-plane analysis for our $w$CDM and $w_{0}w_{a}$CDM models was also trialled, by importance-weighting our posteriors against an s3-plane focusing statistic. This weighting had little effect on the $\Omega_{m}$--$w$ posterior in $w$CDM, but shifted $w_{a}$ more noticeably in favour of $w_{a}=0$ in the $w_{0}w_{a}$CDM model. No model yet includes an s3-focusing term directly in the likelihood or performs surface brightness reconstruction on that plane, and our weighting did not assess agreement between the de-lensed s3 images; a de-lensing of the MUSE data indicated that the morphology of the background source may not be a good depiction of the data under our well-focused solutions. Further line-of-sight complexity is evidently needed for a successful s3 source reconstruction. This, along with more careful modelling of the mass on the s1-plane, will be addressed in future work toward a full triple-source-plane measurement of the dark energy equation of state.

While a posterior shift in cosmology space under a kinematics constraint can be diagnosed as a breaking of the multi-plane mass-sheet degeneracy of \citet{teodori_approximate_2026}, we caution that our constraint suppresses only the degeneracy \textit{internal} to our assumed deflector profile — it does not constrain an \textit{external} convergence sheet, $\kappa_{\rm ext}$, from line-of-sight mass external to any deflector plane. Estimating $\kappa_{\rm ext}$ is typically done in time-delay cosmography via weighted galaxy number counts or weak lensing \citep[e.g.][]{wells_tdcosmo_2023}, but this has not yet been attempted for compound lensing dark energy constraints, so it remains unknown to what extent an external mass-sheet component from over- or under-dense environments could bias $w$ in practice.

In summary, constraints on the dark energy equation of state from lens models of J0946 are highly complementary to existing probes once combined with deflector stellar kinematics. Our $w$CDM and $w_{0}w_{a}$CDM posteriors, from simple and perturbed power-law models, are compatible with $\Lambda$CDM, and the latter do not reproduce tensions that can arise between DESI DR2 BAO and other dark energy probes. With further DSPLs already being discovered \citep{euclidcollaboration_euclid_2025c}, and the promise of a ``big-data'' era of strong lensing from surveys like Euclid and Rubin \citep{collett_population_2015}, this remains a promising probe for precision cosmology; we advocate for non-lensing tracers of deflector mass profiles to realise its potential.

\section*{Acknowledgements}
We thank Nada Salama and Jordan Winstanley for helpful comments on our manuscript. We also thank Kim-Vy Tran and Tania Barone for facilitating regular helpful discussions with the AGEL collaboration which have enriched this work. DJB, GFL and KG acknowledge the support of the Australian Research Council Discovery Project DP230101775. This work has also received funding from the European Research Council (ERC) under the European Union's Horizon 2020 research and innovation program (LensEra: grant agreement No 945536). TEC is funded by the Royal Society through a University Research Fellowship. HQ acknowledges the support of the Australian government International Research Training Program Scholarship. This research was undertaken with the assistance of resources from the National Computational Infrastructure (NCI Australia), an NCRIS enabled capability supported by the Australian Government and the Sydney Informatics Hub, a Core Research Facility of the University of Sydney.
%%%%%%%%%%%%%%%%%%%%%%%%%%%%%%%%%%%%%%%%%%%%%%%%%%
\section*{Data Availability}
Research data used in this work are available on request from the corresponding author or can be found from the HST and VLT archives.

% The inclusion of a Data Availability Statement is a requirement for articles published in MNRAS. Data Availability Statements provide a standardised format for readers to understand the availability of data underlying the research results described in the article. The statement may refer to original data generated in the course of the study or to third-party data analysed in the article. The statement should describe and provide means of access, where possible, by linking to the data or providing the required accession numbers for the relevant databases or DOIs.


\bibliographystyle{mnras}
\bibliography{references} % if your bibtex file is called example.bib

%\begin{thebibliography}{99}
%\end{thebibliography}

%%%%%%%%%%%%%%%%%%%%%%%%%%%%%%%%%%%%%%%%%%%%%%%%%%

%%%%%%%%%%%%%%%%% APPENDICES %%%%%%%%%%%%%%%%%%%%%

% If you want to present additional material which would interrupt the flow of the main paper,
% it can be placed in an Appendix which appears after the list of references.

\appendix
\section{Individual lens model discussion}
\label{app:individual_lens_models}

\begin{figure*}
    \centering
    \includegraphics[width=\linewidth]{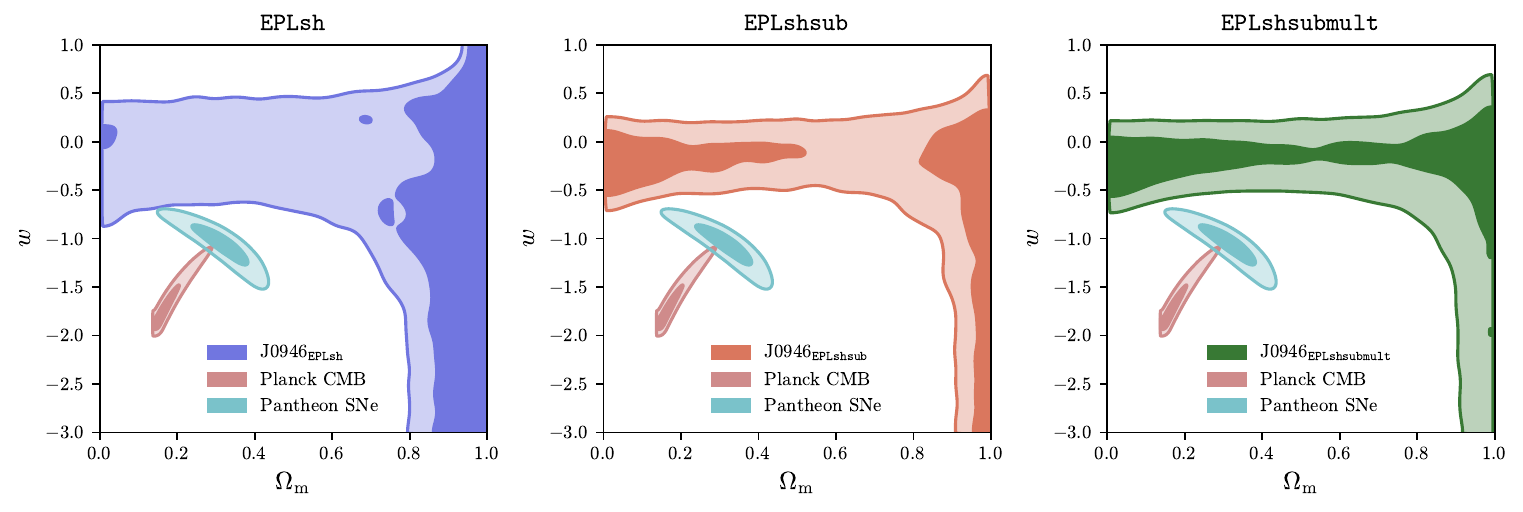}
    %{Figures/w_Om_mass_model_comparison_no_kin.pdf}
    \caption{Posterior distributions in the $w_{0}$--$\Omega_{m}$ plane for our $w$CDM analysis of J0946 with a lensing-only likelihood. Planck CMB and Pantheon SNe Ia constraints are overlaid. All contours denote $1\sigma$ and $2\sigma$ confidence levels.}
    \label{fig:DSPL_mass_model_comparison_no_kin}
\end{figure*}

\begin{figure*}
    \centering
    \includegraphics[width=\linewidth]{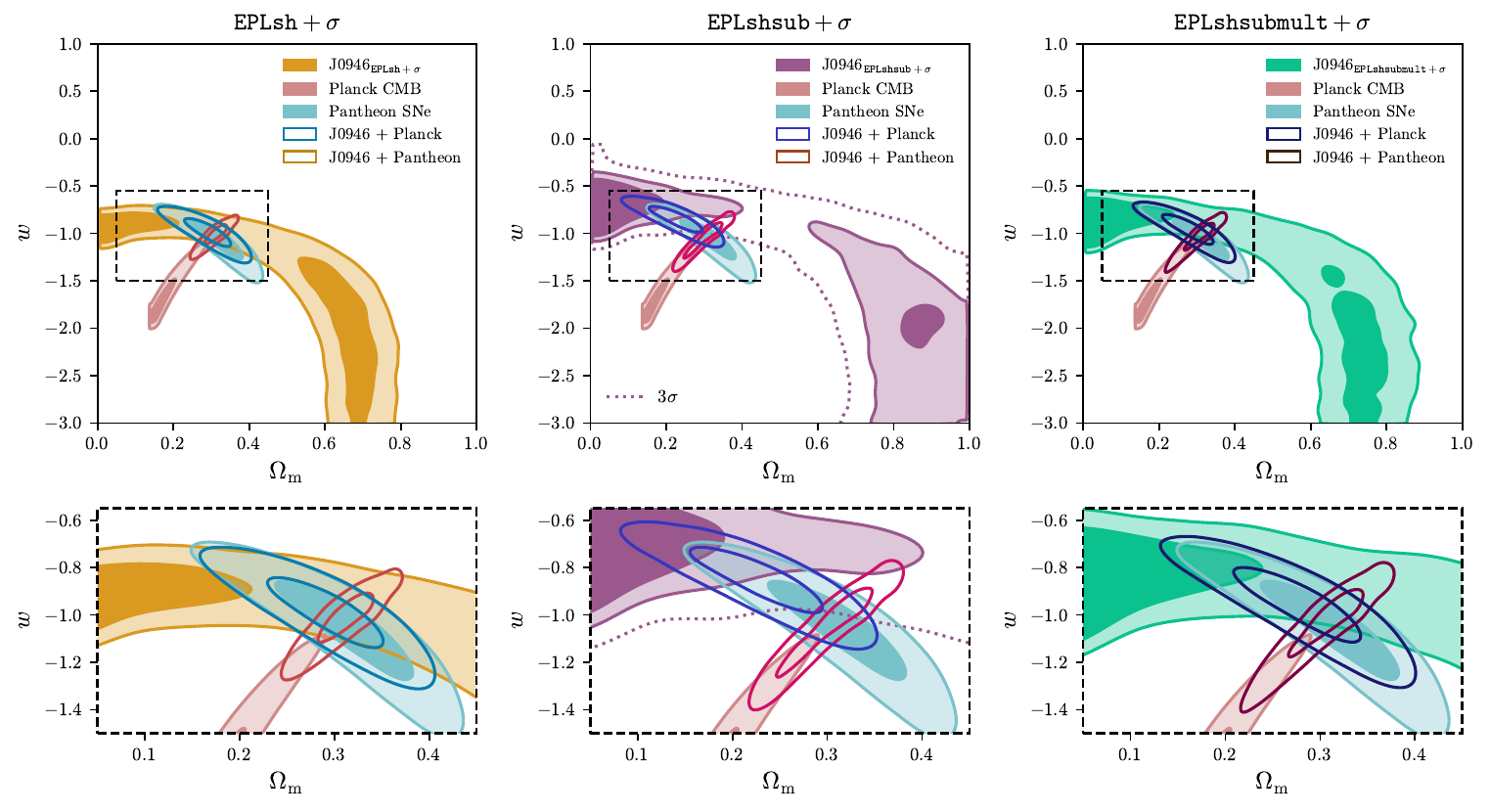}
    \caption{Posterior distributions in the $w_{0}$--$\Omega_{m}$ plane for our $w$CDM J0946$_{*+\sigma}$ analyses. Planck CMB and Pantheon SNe Ia constraints are overlaid and combined with J0946. Bottom panels show a zoom-in of the dashed-box region in the top panels. All contours denote $1\sigma$ and $2\sigma$ confidence levels, with an extra dotted contour denoting the $3\sigma$ contour for the J0946$_{\texttt{EPLshsub}+\sigma}$ model, to clarify that the two distinct islands of the posterior are joined similarly to our J0946$_{\texttt{EPLsh}+\sigma}$ and J0946$_{\texttt{EPLshsubmult}+\sigma}$ cases, but only further into the tail of the posterior distribution.}
    \label{fig:DSPL_mass_model_comparison_kin}
\end{figure*}

\begin{figure*}
    \centering
    \includegraphics[width=\linewidth]{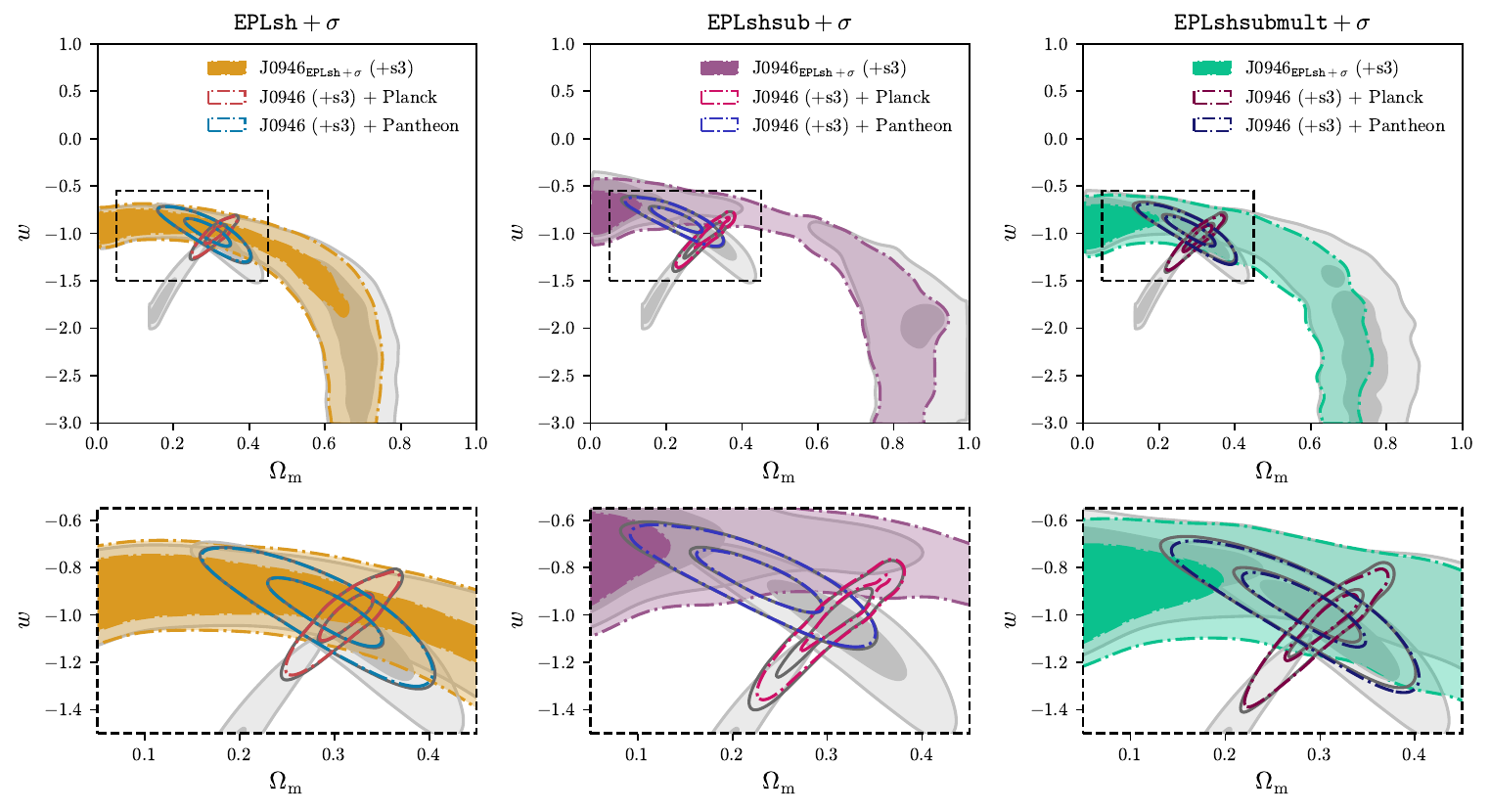}
    \caption{All components of Figure \ref{fig:DSPL_mass_model_comparison_kin}, in greyscale, with s3 importance weighted versions of the J0946, Pantheon+J0946 and DESI+J0946 contours overlaid.}
    \label{fig:DSPL_mass_model_comparison_kin_s3weighted}
\end{figure*}

Here we present the constraints in the $\Omega_{m}$-$w$ plane for each lens model scenario tested, with and without our MUSE kinematics constraint, and briefly discuss the inferred cosmological parameters and properties of the mass model in each individual lens model case. Figure \ref{fig:DSPL_mass_model_comparison_no_kin} shows constraints from imaging data of J0946 only, along with existing Planck CMB \citep{planckcollaboration_planck_2020} and Pantheon SNe Ia \citep{scolnic_complete_2018} constraints. Figure \ref{fig:DSPL_mass_model_comparison_kin} shows constraints from imaging data \textit{and} the kinematics measured by MUSE, and is similarly overlaid with identical Planck and Pantheon constraints, as well as joint posteriors between Planck and J0946 and Pantheon and J0946. Figure \ref{fig:DSPL_mass_model_comparison_kin_s3weighted} shows the components of Figure \ref{fig:DSPL_mass_model_comparison_kin} in greyscale, with coloured and dashdotted contours overlaid to represent how the J0946 and joint posteriors are modified after re-weighting the J0946 posterior with our s3-focusing criterion.

\subsubsection*{Lensing only, without substructure}

By modelling the foreground lens as an EPL with external shear, we find $w = -0.51^{+0.77}_{-1.43}$ and $\Omega_m = 0.77^{+0.18}_{-0.47}$. We infer a deflector on the foreground lens plane with Einstein radius $\vartheta_{E, \rm lens}=1.400\pm 0.001$ and a deflector on the s1 plane with Einstein radius $\vartheta_{E, \rm s1}=0.12\pm 0.01$, with power-law slopes $\gamma_{\rm lens}=1.91\pm 0.01$ and $\gamma_{\rm s1}=2.46^{+0.03}_{-0.07}$ respectively. 

\subsubsection*{Lensing only, with substructure}

Adding a tNFW substructure component to the foreground mass model influences the precision on the inferred cosmology not insignificantly, yielding $w = -0.26^{+0.33}_{-0.86}$ and $\Omega_m = 0.71^{+0.25}_{-0.54}$. This model yields Einstein radii $\vartheta_{E, \rm lens}, \vartheta_{E, \rm s1}=1.396^{+0.001}_{-0.003}, 0.15^{+0.01}_{-0.04}$ and power law slopes $\gamma_{\rm lens}, \gamma_{\rm s1}=1.95^{+0.14}_{-0.01}, 2.37^{+0.09}_{-0.71}$. 

Though the focus of this work is not to address the properties of the inferred perturbing subhalo, we note that our posterior distribution has a substructure with virial mass $\log_{10}\left(M_{200}/M_{\odot}\right)=9.40^{+0.21}_{-0.12}$ and concentration $c_{200}=387^{+219}_{-171}$, which is high but imprecise, and as such is $\lesssim2\sigma$ consistent with dark subhalo simulations. 

\subsubsection*{Lensing only, with substructure and multipoles}

Allowing for non--zero $1^{\text{st}}$-, $3^{\text{rd}}$- and $4^{\text{th}}$-order multipolar moments on the main deflector in addition to substructure has only a modest effect on the inferred cosmology. This model yields cosmological constraints of $w = -0.25^{+0.30}_{-0.75}$ and $\Omega_{m} = 0.68^{+0.27}_{-0.51}$, Einstein radii $\vartheta_{E, \rm lens}, \vartheta_{E, \rm s1}=1.394^{+0.002}_{-0.002}, 0.14\pm 0.03$ and power law slopes $\gamma_{\rm lens}, \gamma_{\rm s1}=2.01^{+0.20}_{-0.06}, 2.06^{+0.33}_{-0.49}$.

The inferred substructure has $\log_{10}\left(M_{200}/M_{\odot}\right)=9.41^{+0.36}_{-0.18}$ and concentration $c_{200}=572.26^{+292.44}_{-304.23}$, which is similarly discrepant versus simulation as our previously described model with no multipoles.

The multipole amplitudes constrained by this model are $A_{\mathcal{M}_1}, A_{\mathcal{M}_3}, A_{\mathcal{M}_4}=0.04^{+0.04}_{-0.01}, 0.005^{+0.003}_{-0.002}, 0.011^{+0.004}_{-0.003}$. These are consistent with previous findings in \citet{enzi_overconcentrated_2024}, excepting the $4^{\rm th}$ order multipolar definition, which is significantly weaker in our model.

\subsubsection*{Lensing and kinematics, without substructure}

When including the kinematics of the foreground lens into the likelihood function, but with all lens model priors kept the same, our fiducial, substructure-less model infers cosmological parameters $w = -1.61^{+0.67}_{-0.89}$ and $\Omega_m = 0.62^{+0.10}_{-0.41}$, Einstein radii $\vartheta_{E, \rm lens}, \vartheta_{E, \rm s1}=1.401\pm0.001, 0.11\pm0.01$ and power law slopes $\gamma_{\rm lens}, \gamma_{\rm s1}=1.88\pm 0.01, 2.46^{+0.03}_{-0.07}$.

\subsubsection*{Lensing and kinematics, with substructure}

Adding substructure whilst including the kinematics constraint yields a slight shift in the cosmological parameters. This model yields $w = -1.16^{+0.49}_{-1.22}$ and $\Omega_m = 0.69^{+0.20}_{-0.65}$. The inferred Einstein radii are $\vartheta_{E, \rm lens}, \vartheta_{E, \rm s1}=1.399^{+0.001}_{-0.001}, 0.14\pm 0.01$ and power law slopes are $\gamma_{\rm lens}, \gamma_{\rm s1}=1.92\pm 0.01, 2.46^{+0.03}_{-0.06}$. 

The inferred substructure is less massive and more concentrated when kinematics data are included. Its virial mass is $\log_{10}\left(M_{200}/M_{\odot}\right)=9.23^{+0.10}_{-0.08}$ and its concentration is $c_{200}=778^{+207}_{-269}$. Solutions consistent with simulations are found between $2-3\sigma$ away from the median.

\subsubsection*{Lensing and kinematics, with substructure and multipoles}

Whilst including kinematics, adding multipoles as well as a substructure shifts cosmology less significantly from the fiducial model than when a substructure is added alone. This model yields $w = -1.47^{+0.62}_{-1.02}$ and $\Omega_m = 0.64^{+0.14}_{-0.55}$. The Einstein radii are constrained to $\vartheta_{E, \rm lens}, \vartheta_{E, \rm s1}=1.395\pm0.002, 0.11\pm 0.01$, and the power law slopes to $\gamma_{\rm lens}, \gamma_{\rm s1}=1.88\pm 0.01, 2.44^{+0.05}_{-0.10}$.

The parameters of the inferred substructure in this case have larger uncertainties, with $\log_{10}\left(M_{200}/M_{\odot}\right)=9.22^{+0.56}_{-0.14}$ and concentration $c_{200}=605.17^{+337.66}_{-369.21}$, which is therefore again $<2\sigma$ in line with dark matter subhalo simulations.

The multipole amplitudes constrained by this model are $A_{\mathcal{M}_1}, A_{\mathcal{M}_3}, A_{\mathcal{M}_4}=0.05\pm 0.01, 0.004\pm 0.002, 0.008\pm 0.002$. These are very similar to the inferred amplitudes without kinematics included.

\section{1D cosmological parameter posteriors in $w$CDM}
\label{sec:ridge_plots_wCDM}
\begin{figure}
    \centering
    \includegraphics[width=\linewidth]{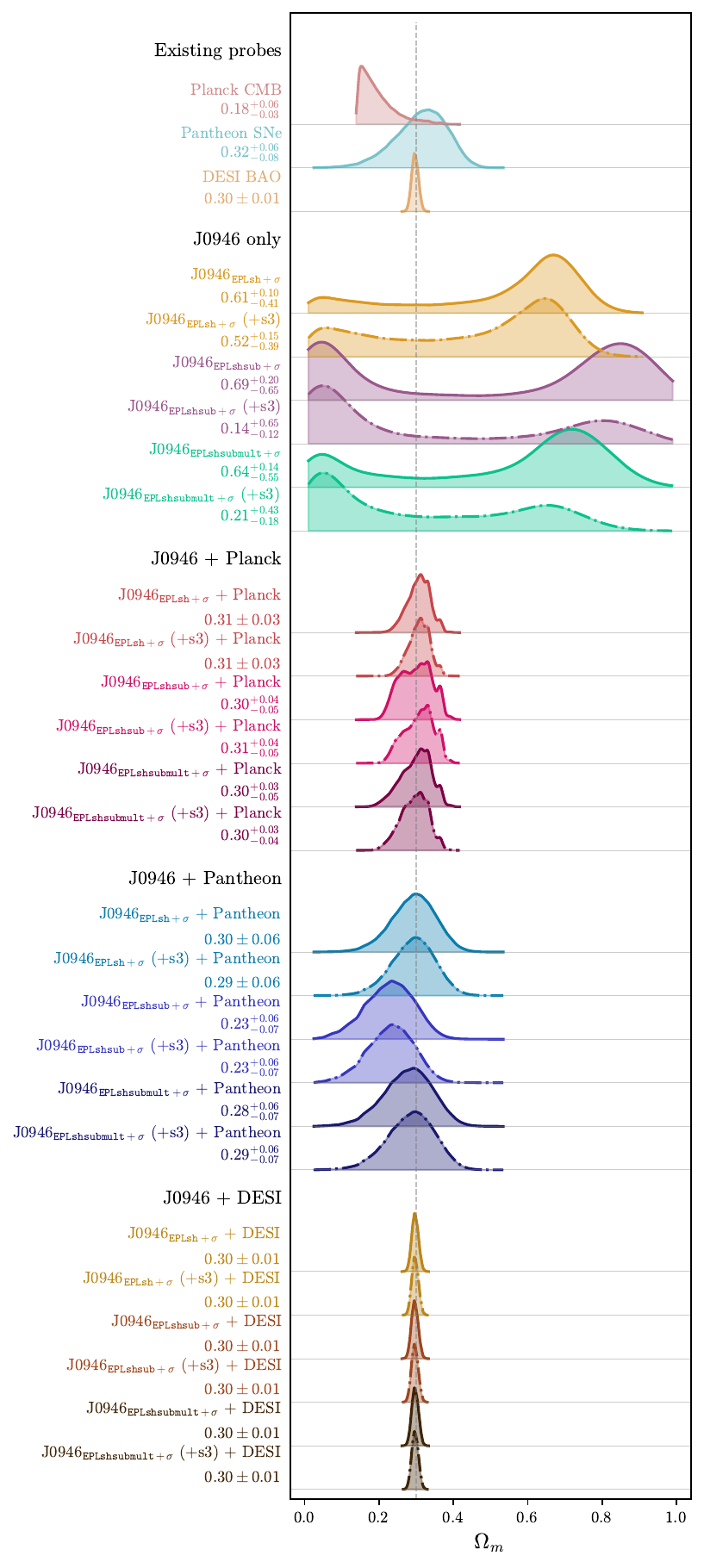}
    \caption{Planck, Pantheon and DESI posteriors, all of our J0946$_{*+\sigma}$ posteriors, and joint posteriors of J0946$_{*+\sigma}$ with each external probe of $\Omega_{m}$ in a $w$CDM cosmology.}
    \label{fig:Om_wCDM_ridge_plot}
\end{figure}

\begin{figure}
    \centering
    \includegraphics[width=\linewidth]{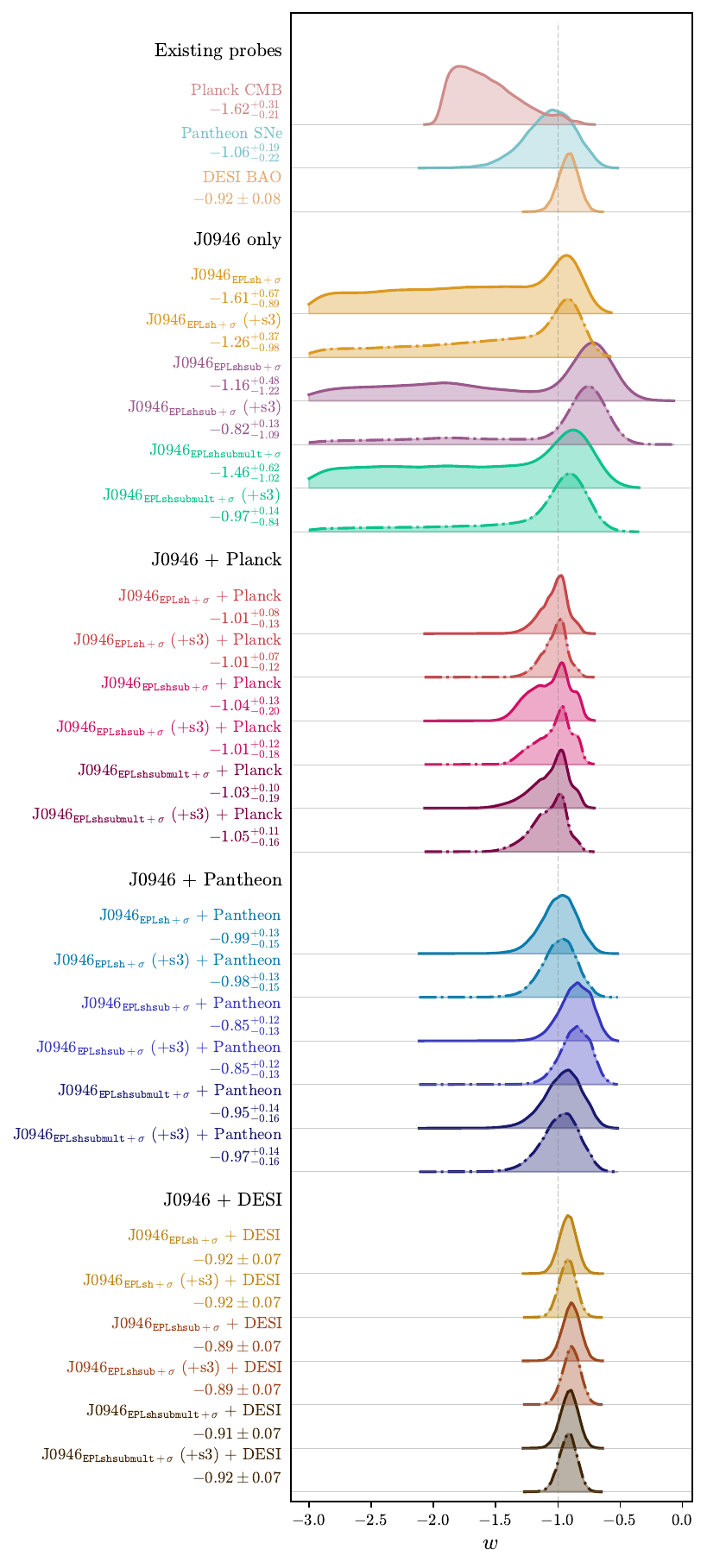}
    \caption{Likewise as in Figure \ref{fig:Om_wCDM_ridge_plot}, for $w$ in $w$CDM.}
    \label{fig:w_wCDM_ridge_plot}
\end{figure}

In Figures \ref{fig:Om_wCDM_ridge_plot} and \ref{fig:w_wCDM_ridge_plot}, we show the shapes of our $\Omega_{m}$ and $w$ posteriors, respectively, for all of our J0946$_{*+\sigma}$ models performed in this work. These are projected into one dimension, and weighted against Planck CMB \citep{planckcollaboration_planck_2020}, Pantheon SNe Ia \citep{scolnic_complete_2018} and also DESI BAO \citep{desicollaboration_desi_2025} constraints.

\section{1D cosmological parameter posteriors in $w_{0}w_{a}$CDM}
\label{sec:ridge_plots_w0waCDM}
In Figures \ref{fig:Om_wCDM_ridge_plot} and \ref{fig:w_wCDM_ridge_plot}, we show the shapes of our $\Omega_{m}$, $w_{0}$ and $w_{a}$ posteriors, respectively, for our J0946$_{\texttt{EPLsh}+\sigma}$ $w_{0}w_{a}$CDM model, showed unweighted and weighted against Planck CMB \citep{planckcollaboration_planck_2020}, Pantheon SNe Ia \citep{scolnic_complete_2018} and also DESI BAO \citep{desicollaboration_desi_2025} constraints.
\begin{figure}
    \centering
    \includegraphics[width=\linewidth]{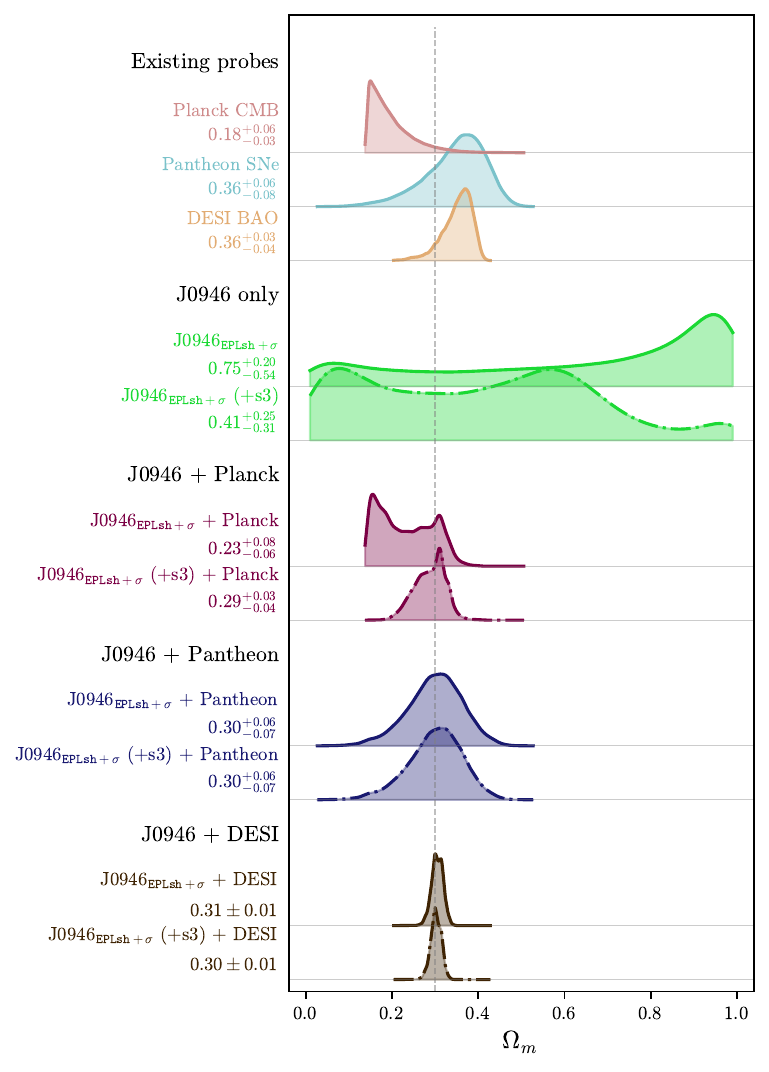}
    \caption{Planck, Pantheon and DESI posteriors, our J0946$_{\texttt{EPLsh}+\sigma}$ posteriors, and joint posteriors of J0946$_{\texttt{EPLsh}+\sigma}$ with each external probe of $\Omega_{m}$ in a $w_{0}w_{a}$CDM cosmology.}
    \label{fig:Om_w0waCDM_ridge_plot}
\end{figure}
\begin{figure}
    \centering
    \includegraphics[width=\linewidth]{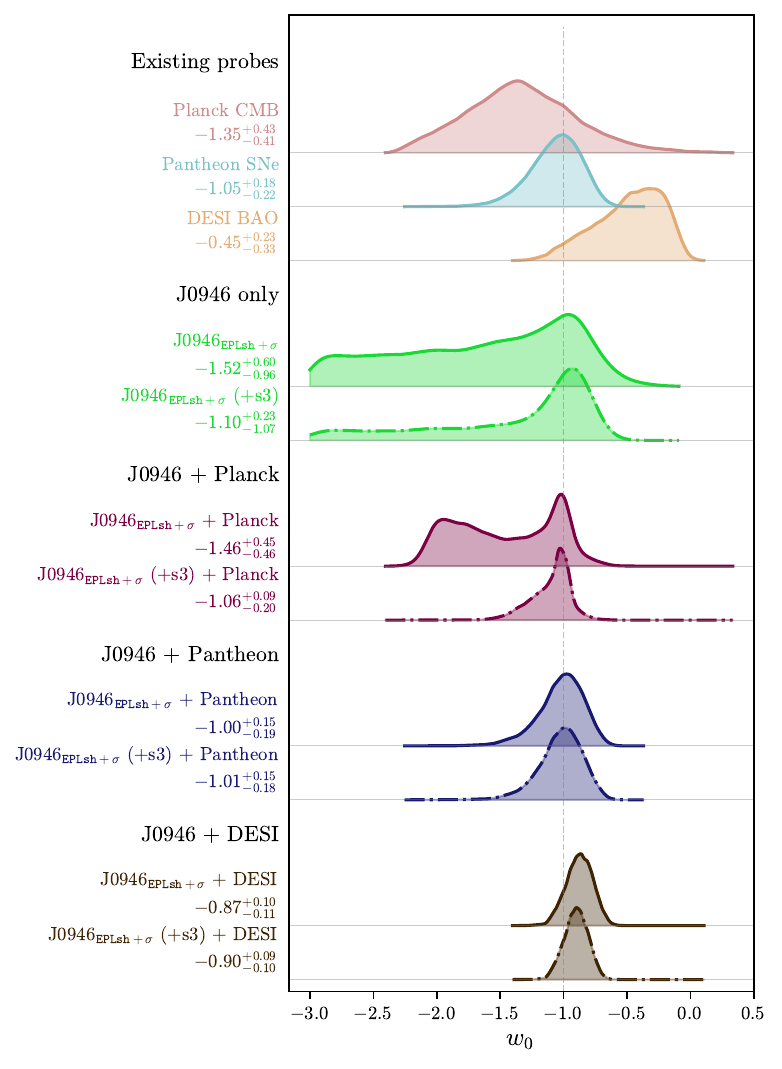}
    \caption{Likewise as in Figure \ref{fig:Om_wCDM_ridge_plot}, for $w_{0}$ in $w_{0}w_{a}$CDM.}
    \label{fig:w0_w0waCDM_ridge_plot}
\end{figure}
\begin{figure}
    \centering
    \includegraphics[width=\linewidth]{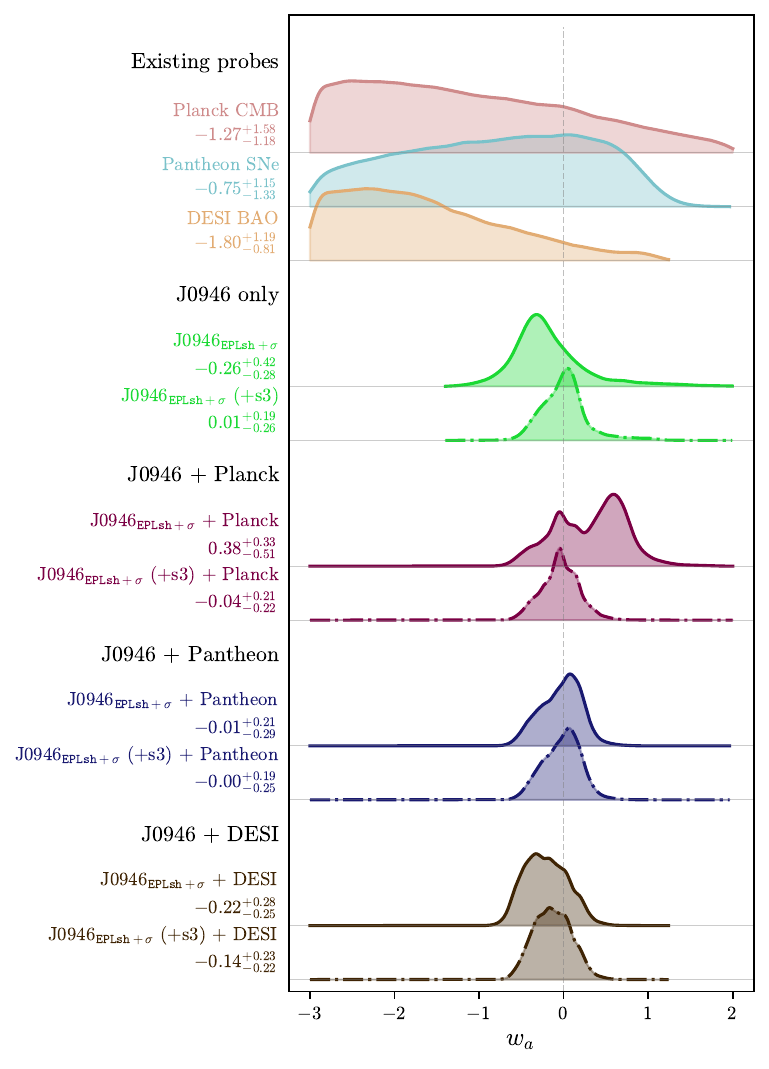}
    \caption{Likewise as in Figure \ref{fig:Om_wCDM_ridge_plot}, for $w_{a}$ in $w_{0}w_{a}$CDM.}
    \label{fig:wa_w0waCDM_ridge_plot}
\end{figure}

%%%%%%%%%%%%%%%%%%%%%%%%%%%%%%%%%%%%%%%%%%%%%%%%%%

% Don't change these lines
\bsp	% typesetting comment
\label{lastpage}
\end{document}